\documentclass[12pt,peerreview,onecolumn]{IEEEtranTCOM}
\pdfoutput=1
\usepackage{graphicx}
\usepackage{amssymb}
\usepackage{subfigure}
\usepackage{cite}
\usepackage[cmex10]{amsmath}
\usepackage{color}
\usepackage{cite}
\usepackage{epsfig}
\usepackage{caption}
\usepackage{cite}
\usepackage{url}
\usepackage{enumerate}
\usepackage[normalem]{ulem}
\usepackage{subfloat}
\usepackage{floatrow}
\usepackage{fixltx2e}

\newcommand{\M}[1]{\mathrm{#1}}

\newtheorem{definition}{Definition}
\newtheorem{lemma}{Lemma}
\newtheorem{proposition}{Proposition}
\newtheorem{remark}{Remark}
\newtheorem{theorem}{Theorem}

\ifCLASSOPTIONpeerreview
\renewcommand{\baselinestretch}{1.87}
\fi

\begin{document}
\let\svthefootnote\thefootnote
\title{Superimposed Signaling Inspired Channel Estimation in Full-Duplex Systems}
\author{
    \IEEEauthorblockN{Abbas Koohian,~\IEEEmembership{Student Member,~IEEE}, Hani Mehrpouyan,~\IEEEmembership{Member,~IEEE}, Ali A. Nasir,~\IEEEmembership{Member,~IEEE}, Salman Durrani,~\IEEEmembership{Senior Member,~IEEE}, Mohammad Azarbad, Steven D. Blostein,~\IEEEmembership{Senior Member,~IEEE}.}
   }
\maketitle
\maketitle

\begin{abstract}
Residual self-interference (SI) cancellation in the digital baseband is an important problem in full-duplex (FD) communication systems. In this paper, we propose a new technique for estimating the SI and communication channels in a FD communication system, which is inspired from superimposed signalling. In our proposed technique, we add a constant real number to each constellation point of a conventional modulation constellation to yield asymmetric shifted modulation constellations with respect to the origin.  We show mathematically that such constellations can be used for bandwidth efficient channel estimation without ambiguity. We propose an expectation maximization (EM) estimator for use with the asymmetric shifted modulation constellations. We derive a closed-form lower bound for the mean square error (MSE) of the channel estimation error, which allows us to find the minimum shift energy needed for accurate channel estimation in a given FD communication system. The simulation results show that the proposed technique outperforms the data-aided channel estimation method, under the condition that the pilots use the same extra energy as the shift, both in terms of MSE of channel estimation error and bit error rate. The proposed technique is also robust to an increasing power of the SI signal. \let\thefootnote\relax\footnote{Abbas Koohian and Salman Durrani are with Research School of Engineering, Australian National University, Canberra, Australia (email: \{abbas.koohian, salman.durrani\}@anu.edu.au). Hani Mehrpouyan is with the Department of Electrical and Computer Engineering, Boise State University, Idaho, USA (email: hani.mehr@ieee.org). Ali A. Nasir is with Department of Electrical Engineering, King Fahd University of Petroleum and Minerals, Dhahran, Saudi Arabia (email: anasir@kfupm.edu.sa). Mohammad Azarbad is with the Department of Mathematics and Statistics, Shahid Chamran University, Ahvaz, Iran (email: m-azarbad@stu.scu.ac.ir). Steven D. Blostein is with the Department of Electrical and Computer Engineering, Queen's University, Kingston, Ontario, Canada (email:steven.blostein@queensu.ca).}
\end{abstract}

\ifCLASSOPTIONpeerreview
\vspace{-0.5 cm}
\fi
\newpage
\setcounter{page}{2}
\renewcommand{\baselinestretch}{2}
\section{Introduction}\label{sec:intro}
\textit{\underline{Background:}} Full-duplex (FD) communication, allowing devices to transmit and receive over the same temporal and spectral resources, is a promising mechanism to potentially double the spectral efficiency of future wireless communication systems~\cite{Korpi-2016}. The main practical challenge in realizing FD communication systems is managing the strong self-interference (SI) signal caused by the transmitter antenna at the receiver antenna within the same transceiver~\cite{Zhang-2015,Sabharwal-2014}. This strong self-interference signal has to be suppressed to the receiver noise floor in order to ensure that it does not degrade the system performance. For instance, for small-cells in Long Term Evolution (LTE), the maximum transmit power is typically $23$ dBm ($200$ mW) and the typical noise floor is $-90$ dBm. Ideally, this requires a total of $113$ dB SI cancellation for realizing the full potential of FD systems~\cite{Sim-2016}.

Recently, there has been a lot of interest in SI cancellation techniques for FD systems~\cite{Duarte-2012,Everett-2014,Kaufman-2013,Korpi-2014,Korpi-2014b,Ahmed-15,Nadh-2016}. The SI cancellation techniques in the literature can be divided into two main categories~\cite{Duarte-2012}: (i) \textit{passive} suppression in which the SI signal is suppressed by suitably isolating the transmit and receive antennas~\cite{Duarte-2012,Everett-2014}, and (ii) \textit{active} cancellation which uses knowledge of the SI signal to cancel the interference in either the analog domain (i.e., before the signal passes through the analog-to-digital converter)~\cite{Duarte-2012,Kaufman-2013} and/or the digital domain~\cite{Korpi-2014,Korpi-2014b,Ahmed-15}. Depending upon the design, passive suppression and analog cancellation can provide about $40-60$ dB cancellation in total~\cite{Nadh-2016}. Hence, in practice, the SI is cancelled in multiple stages, beginning with passive suppression and followed by cancellation in the analog and digital domains. In this paper, we focus on the SI after the passive suppression and analog cancellation, termed \textit{residual SI}.

\textit{\underline{Motivation and Related Work:}} The residual SI can still be relatively strong in the baseband digital signal, e.g., for the LTE small-cell example, it can be as high as $50$ dB assuming state-of-the-art passive suppression and analog cancellation provide $60$ dB of the total required SI cancellation of $113$ dB. Thus, accurate digital SI cancellation is required to bring the SI as close to the noise floor as possible. The effectiveness of any digital interference cancellation technique depends strongly on the quality of the available channel estimates for both the SI and desired communication channels~\cite{Kim-2013,Masmoudi-2015}. Typically, the baseband channels are estimated by using \textit{data-aided channel estimation techniques}, where a portion of the data frame is allocated for known training sequences or pilot symbols~\cite{Li-2011,Masmoudi-2015}. In this regard, a maximum-likelihood (ML) approach was proposed in~\cite{Masmoudi-2015} to jointly estimate the residual SI and communication channels by exploiting the known transmitted symbols and both the known pilot and unknown data symbols from the other intended transceiver. Another approach was proposed in~\cite{Masmoudi-2016} where a sub-space based algorithm was developed to jointly estimate the residual SI and communication channels.

Compared to half-duplex (HD) systems, data-aided channel estimation in FD systems can be bandwidth inefficient. This is because, firstly, two channels need to be estimated and, secondly, accurate channel estimation requires a larger number of pilots~\cite{Koohian-2015}. A bandwidth efficient channel estimation technique in HD systems is \textit{superimposed training}, where no explicit time slots are allocated for channel estimation. Instead, a periodic low power training sequence is superimposed with the data symbols at the transmitter before modulation and transmission~\cite{Mazzenga-2000,Tugnait-2003}. The downside of this approach is that some power is consumed in superimposed training which could have otherwise been allocated to the data transmission. This lowers the effective signal-to-noise ratio (SNR) for the data symbols and affects the bit error rate (BER) at the receiver. In contrast to data-aided and superimposed training based channel estimation techniques, \textit{blind techniques} avoid the use of pilots altogether by exploiting statistical and other properties of the transmitted signal~\cite{Carvalho-2004,Abdallah-2013,Abdallah-2012,Zhang-2013,Zhu-2009}. However, blind estimators can only estimate the channel up to a scaling factor and cannot recover the channel phases~\cite{Carvalho-2004}. The necessary and sufficient conditions for ambiguity-free blind estimation can be determined using identifiability analysis, which determines whether a parameter can be uniquely estimated without any ambiguity~\cite{Carvalho-2004,Lehmann-1998,Tong-1998,Meraim-1997,Koohian-2015}. To the best of our knowledge, bandwidth efficient and accurate channel estimation methods for FD systems is still an important open area of research.

\textit{\underline{Paper Contributions:}} In this paper, we consider the problem of bandwidth efficient channel estimation in a single-input single-output (SISO) FD communication system. We propose a new technique for channel estimation and residual SI cancellation in FD systems. Our approach draws inspiration from (i) blind channel estimation techniques in that we examine the condition for identifiability of channel parameters in FD systems and (ii) superimposed signalling in that we superimpose (i.e., add) a constant real number to each constellation point of the modulation constellation. However, the nature and objective of the superimposed signal in our proposed technique is different to that in superimposed signalling. In superimposed signalling, the superimposed signal is typically a periodic training sequence that is added to the data signal after the data symbols are modulated. Hence, the additional power of the superimposed signal is only used for channel estimation. In our proposed technique, the superimposed signal is a constant (non-random) signal and the objective is to shift the modulation constellation away from the origin, which we exploit for estimating the SI and communications channels without ambiguity. In addition, the additional power of the superimposed signal is used for both modulating the data symbols and channel estimation, which does not reduce the effective SNR as in superimposed signalling. The novel contributions are as follows:
\begin{itemize}
\item We derive the condition for identifiability of channel parameters in a FD system (cf. Theorem~\ref{theorem:identifiability}) and show that symmetric modulation constellations with respect to the origin cannot be used for ambiguity-free channel estimation in a FD system. Based on Theorem~\ref{theorem:identifiability}, our proposed technique is able to resolve the inherent ambiguity of blind channel estimation in FD communication via shifting the modulation constellation away from origin.

\item Using the proposed technique, we derive a computationally efficient expectation maximization (EM) estimator for simultaneous estimation of both SI and communication channels. We derive a lower bound for the channel estimation error, which depends on the energy used for shifting the modulation constellations, and use it to find the minimum signal energy needed for accurate channel estimation in a given FD communication system.

\item We use simulations to compare the performance of the proposed technique against that of the data-aided channel estimation method, under the condition that the pilots use the same extra power as the shift. Our results show that the proposed technique performs better than the data-aided channel estimation method both in terms of the minimum mean square error (MSE) of channel estimation and BER. In addition, the proposed technique is robust to an increasing SI power.
\end{itemize}

\textit{\underline{Notation and Paper Organization:}} The following notation is used in this paper. Capital letters are used for random variables and lower case letters are used for their realizations. $f_Y(y)$ denotes the probability density function (PDF) of random variable $Y$. $\mathbb{E}_{Y}[\cdot]$ denotes the expectation with respect to the random variable $Y$. $p_X(x)$ denotes the probability mass function (PMF) of a discrete random variable $X$ and $P(X=a)$ is the probability of the discrete random variable $X$ taking the value $a$. $\mathcal{CN}(\mu,\sigma^2)$ denotes a complex Gaussian distribution with mean $\mu$ and variance $\sigma^2$. Bold face capital letters, e.g., $\mathbf{Y}$ are used for random vectors and bold face lower case letters, e.g., $\mathbf{y}$, are used for their realizations. Capital letters in upright Roman font, e.g., $\mathrm{G}$, are used for matrices. Lower case letters in upright Roman font, e.g., $\mathrm{g}$, are used for functions. $\M{I}_N$ represents the $N\times N$ identity matrix. $[\cdot]^T$ denotes vector and matrix transpose. $\mathrm{j}\triangleq \sqrt{-1}$, and the real and imaginary parts of a complex quantity are represented by $\Re\{\cdot\}$ and $\Im\{\cdot\}$, respectively. $z^*$ and $|z|$ indicate scalar complex conjugate and the absolute value of complex number $z$, respectively. Finally, $\det(\cdot)$ is the determinant operator.

This paper is organized as follows. The system model is presented in Section~\ref{sec:sysmodel}. The channel estimation problem and the proposed technique are formulated in Section~\ref{sec:channel}. The EM estimator and the lower bound on the channel estimator error are derived in Section~\ref{sec:proposed}. The performance of the proposed technique is assessed in Section~\ref{sec:results}. Finally, conclusions are presented in Section~\ref{sec:conc}.

\section{System Model}\label{sec:sysmodel}

Consider the channel estimation problem for a SISO FD communication system between two nodes $a$ and $b$, as illustrated in Fig.~\ref{fig:SysMod}. Transceiver nodes $a$ and $b$ are assumed to have passive suppression and analog cancellation stages and we only consider the digital cancellation to remove the residual SI, i.e., the SI, which is still present after the passive suppression and analog cancellation. We consider the received signal available at the output of the analog to digital converter (ADC). The received signal at node $a$ is given by\footnote{Note that the system model for FD communication in~\eqref{eq:nodea} is applicable to per subcarrier communication in orthogonal frequency division multiplexing (OFDM) FD communication~\cite{Duarate-2010,Duarte-2012,Sabharwal-2014}.}
\ifCLASSOPTIONpeerreview
\begin{figure}[t!]
  \centering
  \scalebox{0.8}{\includegraphics{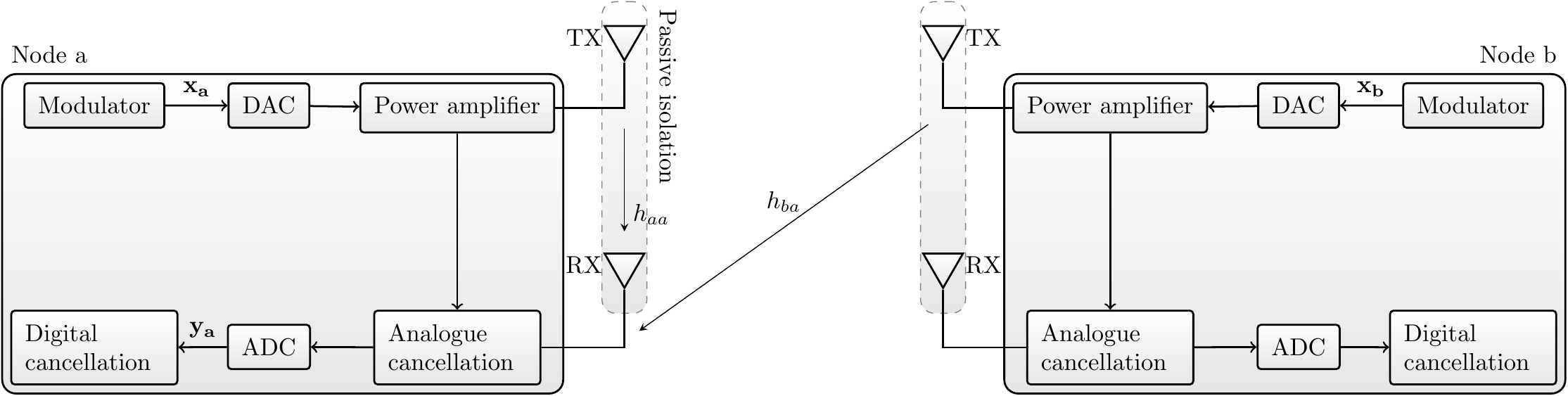}}
  \centering
  \caption{Illustration of full duplex communication between two transceivers, each with a single transmit and a single receive antenna. ADC = analog to digital converter, DAC = digital to analog converter. TX = transmit. RX = receive.}\label{fig:SysMod}
\end{figure}
\else
\begin{figure}[t!]
  \centering
  \scalebox{0.5}{\includegraphics{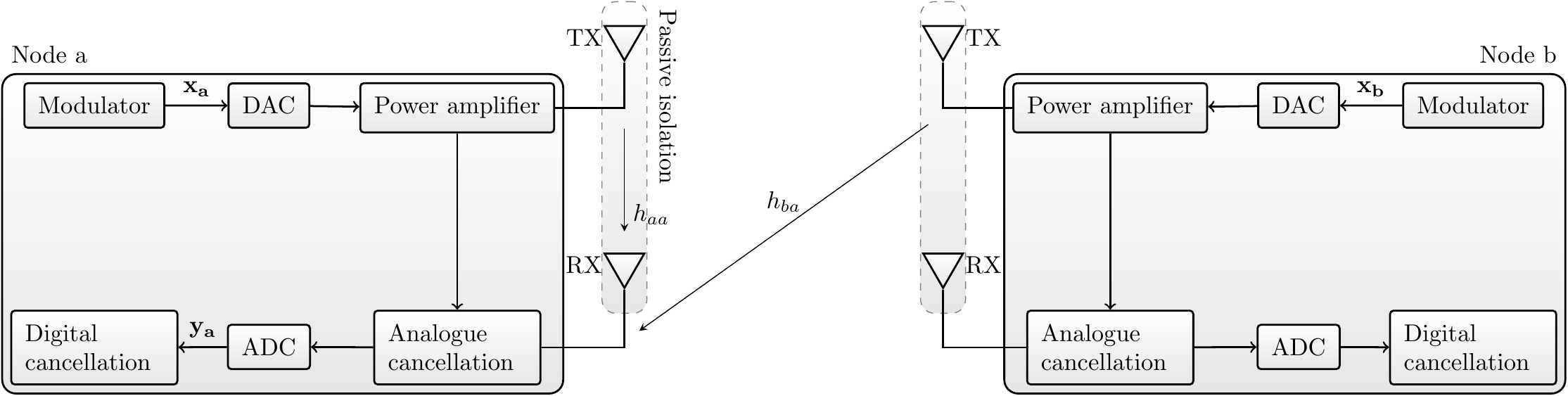}}
  \centering
  \caption{Illustration of full duplex communication between two transceivers, each with a single transmit and a single receive antenna. The single antenna at each node is shown separately for the transmission and reception for ease of illustration. ADC = analog to digital converter, DAC = digital to analog converter. TX = transmit. RX = receive.}\label{fig:SysMod}
\end{figure}
\fi
\begin{align}\label{eq:nodea}
  \mathbf{y_a} = h_{aa}\mathbf{x_{a}}+h_{ba}\mathbf{x_b}+\mathbf{w_a},
\end{align}

\noindent where $\mathbf{x_{a}}\triangleq\left[x_{a_1},\cdots, x_{a_N}\right]^T$, $\mathbf{x_{b}}\triangleq\left[x_{b_1},\cdots, x_{b_N}\right]^T$  are the $N \times 1$ vectors of transmitted symbols from nodes $a$ and $b$, respectively, $\mathbf{y_a}\triangleq \left[y_{a_1},\cdots, y_{a_N}\right]^T $ is the $N \times 1$ vector of observations, $\mathbf{w_a}$ is the noise vector, which is modeled by $N$ independent Gaussian random variables, i.e., $f_{\mathbf{W_a}}(\mathbf{w_a}) = \mathcal{CN}(\mathbf{0},\sigma^2 \M{I}_N)$, and $h_{aa}$, $h_{ba}$ are the residual SI and communication channel gains, respectively. Furthermore, we model $h_{aa}$ and $h_{ba}$ as independent random variables that are constant over one frame of data and change independently from frame to frame~\cite{Kim-2013}.

\textit{\underline{Modulation Assumptions and Definitions:}} In this paper, we assume that the transmitted symbols are all equiprobable and call the set $\mathcal{A}\triangleq \{x_1,x_2,...,x_M\}$, which contains an alphabet of $M$ constellation points, \emph{a modulation set}. Let $\mathcal{K} \triangleq \{1,\cdots,M\}$ denote set of indices of the constellation points.

We define $E$ as the average symbol energy of a given constellation, i.e.,
\begin{align}\label{eq:E}
E\triangleq \mathbb{E}_{X_k}[|x_k|^2]=\frac{\sum_{k=1}^M |x_k|^2}{M},
\end{align}

\noindent where $x_k \in \mathcal{A}$. Note that the average symbol energy can be related to the average bit energy as $E_b \triangleq E/\log_2(M)$.


\section{Channel Estimation for FD systems}\label{sec:channel}
In this section, we first formulate the blind channel estimation problem for the FD system considered in Section~\ref{sec:sysmodel}. Based on this formulation, we present a theorem which provides the necessary and sufficient condition for ambiguity-free channel estimation. Finally, we discuss the proposed technique to resolve the ambiguity problem.

\subsection{Problem Formulation}\label{sec:probfor}

Without loss of generality, we consider the problem of baseband channel estimation at node $a$ only (similar results apply at node $b$). In formulating the problem, we make the following assumptions: (i) the transmitter is aware of its own signal, i.e., $\mathbf{x_a}$ is known at node $a$, which is a commonly adopted assumption in the literature~\cite{Sabharwal-2014,Duarte-2012}, (ii) the interference channel $h_{aa}$, and the communication channel $h_{ba}$ are unknown deterministic parameters, (iii) the transmit symbol from node $b$ is modelled using a discrete random distribution, and (iv) we observe $N$ independent received symbols.

The blind channel estimation problem requires the knowledge of the joint probability density function (PDF) of all observations, which is derived from the conditional PDF of a single observation. Given the system model in~\eqref{eq:nodea}, the conditional PDF of a single observation is given by
\ifCLASSOPTIONpeerreview
\begin{equation}\label{eq:PDFOneObservation}
f_{Y_{a_i}}(y_{a_i}|x_{b_i};h_{aa},h_{ba})=\frac{1}{\pi \sigma^2} \exp \left(\frac{-1}{\sigma^2} |y_{a_i}-h_{ba} x_{b_i}-h_{aa}x_{a_i}|^2 \right),
\end{equation}
\else
\begin{align}\label{eq:PDFOneObservation}
f_{Y_{a_i}}(y_{a_i}|x_{b_i};&h_{aa},h_{ba})= \nonumber\\
&\frac{1}{\pi \sigma^2} \exp \left(\frac{-1}{\sigma^2} |y_{a_i}-h_{ba} x_{b_i}-h_{aa}x_{a_i}|^2 \right),
\end{align}
\fi

\noindent where $i \in \mathcal{I}\triangleq \{1,\cdots,N\}$, $y_{a_i}$ is the $i^{th}$ received symbol, and $x_{a_i}$ and $x_{b_i}$ are the $i^{th}$ transmitted symbols from nodes $a$ and $b$, respectively.

The  marginal PDF of a single observation is then found by multiplying~\eqref{eq:PDFOneObservation} by the uniform distribution $p_{X_{b_i}}(x_{b_i})=\frac{1}{M}\mathbb{I}_{\{\mathcal{A}\}}(x_{b_i})$, and summing the results over all the possible values of $x_{b_i}$, where, $\mathbb{I}_{\{\mathcal{A}\}}(x) =1$ if $x\in \mathcal{A}$ and $0$ otherwise. Therefore, we have
\ifCLASSOPTIONpeerreview
\begin{align}\label{eq:Marginal}
f_{Y_{a_i}}(y_{a_i};h_{aa},h_{ba}) &=\sum_{\forall x_{b_i}} f_{Y_{a_i}}(y_{a_i}|x_{b_i};h_{aa},h_{ba})p_{X_{b_i}}(x_{b_i})\nonumber\\
&=\frac{1}{M\pi\sigma^2} \sum_{\forall x_{b_i}}\exp \left(\frac{-1}{\sigma^2} |y_{a_i}-h_{ba} x_{b_i}-h_{aa}x_{a_i}|^2\right)\mathbb{I}_{\{\mathcal{A}\}}(x_{b_i})& \nonumber\\
&=\frac{1}{M\pi\sigma^2} \sum_{k=1}^M\exp \left(\frac{-1}{\sigma^2} |y_{a_i}-h_{ba} x_k-h_{aa}x_{a_i}|^2\right),
\end{align}
\else
\begin{align}\label{eq:Marginal}
&f_{Y_{a_i}}(y_{a_i};h_{aa},h_{ba}) = \sum_{\forall x_{b_i}} f_{Y_{a_i}}(y_{a_i}|x_{b_i};h_{aa},h_{ba})p_{X_{b_i}}(x_{b_i})\nonumber\\
&=\frac{1}{M\pi\sigma^2} \sum_{\forall x_{b_i}}\exp \left(\frac{-1}{\sigma^2} |y_{a_i}-h_{ba} x_{b_i}-h_{aa}x_{a_i}|^2\right)\mathbb{I}_{\{\mathcal{A}\}}(x_{b_i})&& \nonumber\\
&=\frac{1}{M\pi\sigma^2} \sum_{k=1}^M\exp \left(\frac{-1}{\sigma^2} |y_{a_i}-h_{ba} x_k-h_{aa}x_{a_i}|^2\right),
\end{align}
\fi

\noindent where the last step follows from the fact that $\mathbb{I}_{\{\mathcal{A}\}}(x_{b_i})=1$ if and only if $x_{b_i}=x_k$, where $x_k \in \mathcal{A}$. Finally, since the transmitted symbols are assumed independent, and we observe $N$ independent observations, the joint PDF of all the observations is given by
\begin{align}\label{eq:jointpdf1}
f_{\mathbf{Y_a}}(\mathbf{y_a};h_{aa},h_{ba})&=\prod_{i=1}^N f_{Y_{a_i}}(y_{a_i};h_{aa},h_{ba}) \nonumber \\
&= \left(\frac{1}{M\pi \sigma^2}\right)^N \prod_{i=1}^N \sum_{k=1}^M \exp \left(\frac{-1}{\sigma^2} |y_{a_i}-h_{ba} x_k-h_{aa}x_{a_i}|^2\right).&
\end{align}

\noindent where we substitute the value of $f_{Y_{a_i}}(y_{a_i};h_{aa},h_{ba})$ from \eqref{eq:Marginal}.

Using~\eqref{eq:jointpdf1}, we can state the channel estimation problem as shown in the proposition below.

\begin{proposition}\label{prop:blindestimation}
The blind maximum likelihood (ML) channel estimation problem in a SISO FD system is given by
\begin{align}\label{eq:channelestimationproblem}
 \arg \max_{h_{aa},h_{ba}}f_{\mathbf{Y_a}}(\mathbf{y_a};h_{aa},h_{ba}),
\end{align}

\noindent where $f_{\mathbf{Y_a}}(\mathbf{y_a};h_{aa},h_{ba})$ is given by~\eqref{eq:jointpdf1}.
\end{proposition}


In the next subsection, we show that \eqref{eq:channelestimationproblem} does not have a unique solution if modulation sets which are symmetric around the origin are used.

\subsection{Identifiability Analysis}\label{sec:ambiguity}

In this subsection, we present the identifiability analysis for the blind channel estimation problem in \eqref{eq:channelestimationproblem}, which allows us to determine when ambiguity-free channel estimation is possible. For ease of analysis, we first define $\boldsymbol{\theta}\triangleq [h_{aa},h_{ba}]$ and rewrite~\eqref{eq:jointpdf1} as
\ifCLASSOPTIONpeerreview
\begin{align}\label{eq:jointpdfrewritten}
f_{\mathbf{Y_a}}(\mathbf{y_a};\boldsymbol{\theta})= \left(\frac{1}{M\pi \sigma^2}\right)^N \prod_{i=1}^N \sum_{k=1}^M \exp \left(\frac{-1}{\sigma^2} |y_{a_i}-\boldsymbol{\theta}(1)x_{a_i}-\boldsymbol{\theta}(2)x_{k}|^2\right),
\end{align}
\else
\begin{align}\label{eq:jointpdfrewritten}
f_{\mathbf{Y_a}}(\mathbf{y_a}&;\boldsymbol{\theta})= \left(\frac{1}{M\pi \sigma^2}\right)^N \nonumber \\ &\times \prod_{i=1}^N \sum_{k=1}^M \exp \left(\frac{-1}{\sigma^2} |y_{a_i}-\boldsymbol{\theta}(1)x_{a_i}-\boldsymbol{\theta}(2)x_{k}|^2\right),
\end{align}
\fi

\noindent where $\boldsymbol{\theta}(1)$ and $\boldsymbol{\theta}(2)$ represent the first and second elements of $\boldsymbol{\theta}$.

We start the identifiability analysis by presenting the following definition and remark:
\begin{definition}\label{def:ident}~\cite[Definition 5.2]{Lehmann-1998}
If $\mathbf{Y}$ is a random vector distributed according to $f_\mathbf{Y}(\mathbf{y};\boldsymbol{\theta})$, then $\boldsymbol{\theta}$ is said to be unidentifiable on the basis of $\mathbf{y}$, if $\forall\mathbf{y}$ there exists $\boldsymbol{\theta} \neq \boldsymbol{\theta}'$ for which $f_\mathbf{Y}(\mathbf{y};\boldsymbol{\theta})=f_\mathbf{Y}(\mathbf{y};\boldsymbol{\theta}')$.
\end{definition}

\begin{remark}
Definition~\ref{def:ident} states that $\boldsymbol{\theta}$ and $\boldsymbol{\theta}'$ ($\boldsymbol{\theta} \neq \boldsymbol{\theta}'$) cannot be distinguished from a given set of observations if they both result in the same probability density function for the observations. This implies that if $\boldsymbol{\theta}$ is unidentifiable, then it is impossible for any estimator to uniquely determine the value of $\boldsymbol{\theta}$.
\end{remark}


In order to present the main result in this subsection, we first give the definitions of a symmetric modulation constellation~\cite{Pahl-2001} and a bijective function~\cite{loeher-2011}.

\begin{definition}\label{def:symm}
We mathematically define modulation constellation as the graph of the function $\mathrm{f}(x_k)=x_k$, where $x_k \in \mathcal{A}$ $\forall k \in \mathcal{K}$. Then a modulation constellation is  symmetric with respect to the origin if and only if $\mathrm{f}(-x_k)=-\mathrm{f}(x_k)$ $\forall x_k \in \mathcal{A}$~\cite{Pahl-2001}.
\end{definition}

\begin{definition}
Let $\mathcal{C}$ and $\mathcal{D}$ be two sets. A function from $\mathcal{C}$ to $\mathcal{D}$ denoted
$\mathrm{t}: \mathcal{C} \rightarrow \mathcal{D}$ is a bijective function if and only if it is both one-to-one and onto.
\end{definition}

The above definition states that a bijective function is a function between the elements of two sets, where each element of one set is paired with exactly one element of the other set and there are no unpaired elements. Note that a bijective function from a set to itself is also called a permutation~\cite{loeher-2011}.

In this work, we define and use the bijective function $\mathrm{g}$: $\mathcal{K} \rightarrow \mathcal{K}$, i.e., $\mathrm{g}$ is a one-to-one and onto function on $\mathcal{K}\rightarrow \mathcal{K}$. Using this bijective function, we present the main result as below.
\begin{theorem}\label{theorem:identifiability}
There exists a $\boldsymbol{\theta}' \neq \boldsymbol{\theta}$ for which the joint probability density $f_{\mathbf{Y_a}}(\mathbf{y_a};\boldsymbol{\theta})$ given by~\eqref{eq:jointpdfrewritten} is equal to $f_{\mathbf{Y_a}}(\mathbf{y_a};\boldsymbol{\theta}') \; \forall\mathbf{y}_a$, if and only if there exists a bijective function $\mathrm{g}$: $\mathcal{K} \rightarrow \mathcal{K}$, such that $\frac{x_k}{x_{\mathrm{g}(k)}}=c \; \forall k \in \mathcal{K}$, where $c \neq 1$ is a constant and $|c|=1$, i.e., the modulation constellation is symmetric about the origin.
\end{theorem}

\begin{IEEEproof}
We prove the result in Theorem~\ref{theorem:identifiability} in three steps. First, we assume $\boldsymbol{\theta}' \neq \boldsymbol{\theta}$ for which $f_{\mathbf{Y_a}}(\mathbf{y_a};\boldsymbol{\theta})=f_{\mathbf{Y_a}}(\mathbf{y_a};\boldsymbol{\theta}')$ $\forall\mathbf{y_a}$ exists, and show that it leads to a bijective function $\mathrm{g}$ satisfying $\frac{x_k}{x_{\mathrm{g}(k)}}=c$ $\forall k \in \mathcal{K}$. Then, we assume that a bijective function $\mathrm{g}$ satisfying $\frac{x_k}{x_{\mathrm{g}(k)}}=c$ $\forall k \in \mathcal{K}$ exists, and show that there exists a $\boldsymbol{\theta}' \neq \boldsymbol{\theta}$ for which $f_{\mathbf{Y_a}}(\mathbf{y_a};\boldsymbol{\theta})=f_{\mathbf{Y_a}}(\mathbf{y_a};\boldsymbol{\theta}')$ $\forall\mathbf{y_a}$.
Finally, using Definition~\ref{def:symm}, we show that the condition $\frac{x_k}{x_{\mathrm{g}(k)}}=c$ $\forall k \in \mathcal{K}$ is equivalent to the modulation constellation being symmetric with respect to the origin. The details are in Appendix~\ref{appendix:proof:theorem}.
\end{IEEEproof}

\begin{remark}
From Theorem~\ref{theorem:identifiability}, we can see that since the modulation constellations, such as $M$-QAM, satisfy the definition of symmetric modulation constellations in Definition~\ref{def:symm}, the blind channel estimation problem in~\eqref{eq:channelestimationproblem} does not have a unique solution and suffers from an ambiguity problem.
\end{remark}

\subsection{Proposed Technique}\label{sec:resolving}
In this subsection, we present our proposed technique to resolve the ambiguity problem in~\eqref{eq:channelestimationproblem}.
\begin{figure}[t!]
  \centering
  \scalebox{0.75}{\includegraphics{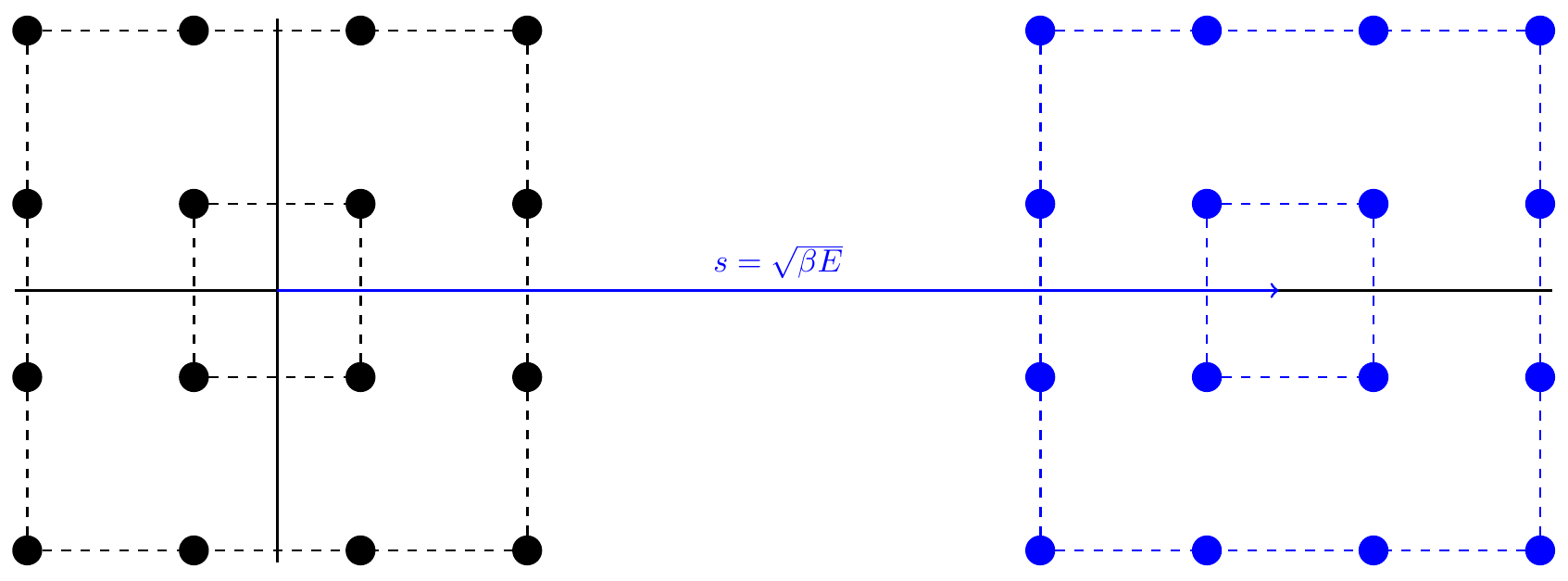}}
  \centering
  \caption{Effect of the proposed technique on the constellation of $16$-QAM. The resulting constellation is shifted along the horizontal axis, i.e., it is asymmetric around the origin.}\label{Fig:QAM}
\end{figure}

The rationale behind the proposed technique comes from the fact that Theorem~\ref{theorem:identifiability} shows that symmetry of the modulation constellation with respect to the origin is the cause of the ambiguity. A simple way to achieve constellation asymmetry\footnote{Note that it may be possible to achieve constellation asymmetry through other means, such as design of irregular modulation constellations. The optimum design of such modulation constellations is outside the scope of this work.} is to add a constant $s$ to each element of $\mathcal{A}$. The resultant asymmetric shifted modulation constellation is formally defined as follows:

\begin{definition}\label{def:asym}
The asymmetric shifted modulation constellation, $\mathcal{\bar{A}}$, is defined as
\begin{align}\label{e:shiftmod q}
\mathcal{\bar{A}}\triangleq \{x_{k}+s, \quad \forall \; x_{k}\in \mathcal{A},\quad s \in \mathbb{R}^+\},
\end{align}

\noindent where $\mathbb{R}^+$ is the set of positive real numbers.
\end{definition}

In the rest of the paper, we also use $\bar{x}_k= x_k+s$ to denote the $k$th element of $\mathcal{\bar{A}}$.

For illustration, Fig.~\ref{Fig:QAM} shows the effect of the proposed technique on the $16$-QAM constellation. We can see that the resulting modulation constellation is shifted along the horizontal axis, which increases the average energy per symbol of the modulation constellation. \textit{This increase in the average energy per symbol can be justified as follows: in reality it is inevitable to use some extra energy to estimate the unknown channels, whether it is done by pilots or by the proposed technique. In this regard, it is important to note that the smaller the energy used for shifting the modulation constellation, the closer the average energy of the proposed technique is to the ideal scenario where the channels are perfectly known at the receiver and no extra energy is needed for channel estimation}.

For the sake of numerically investigating the problem of smallest possible shift energy, we define $\beta$ as the portion of the average energy per symbol that is allocated to the shift and use the real constant $s\triangleq \sqrt{\beta E}$, where $0<\beta<1$ to shift the symmetric modulation constellation. In this case, the problem of smallest shift energy corresponds to the problem of finding the minimum value of $\beta$. The minimum value of $\beta$ is an indication of how much extra energy is needed compared to the perfect channel knowledge scenario.

In Section~\ref{sec:bounds_analysis}, we derive a lower bound on the estimation error, which allows us to numerically find the minimum value of $\beta$.
\section{EM-based Estimator}\label{sec:proposed}
In this section, we derive an EM estimator to obtain channel estimates in a FD system with asymmetric shifted modulation constellation defined in Definition~\ref{def:asym}. We derive a lower bound on the estimation error of the estimator. Finally, we investigate the complexity of the proposed estimator.

For the sake of notational brevity, we first define
\begin{align}\label{eq:def:phi}
\boldsymbol{\phi} \triangleq [\Re(h_{aa}),\Im(h_{aa}),\Re(h_{ba}),\Im(h_{ba})].
\end{align}

\noindent We can then reformulate the ML problem in~\eqref{eq:channelestimationproblem} as follows
\ifCLASSOPTIONpeerreview
\begin{equation}\label{eq:MLE}
 [\hat{\Re(h_{aa})},\hat{\Im(h_{aa})},\hat{\Re(h_{aa})},\hat{\Im(h_{ba})}]\triangleq \arg \max_{\boldsymbol{\phi}}(\ln f_{\mathbf{Y_a}}(\mathbf{y}_\mathbf{a} ; \boldsymbol\phi)),
\end{equation}
\else
{\small
\begin{equation}\label{eq:MLE}
 [\hat{\Re(h_{aa})},\hat{\Im(h_{aa})},\hat{\Re(h_{aa})},\hat{\Im(h_{ba})}]=\arg \max_{\boldsymbol{\phi}}(\ln f_{\mathbf{Y_a}} (\mathbf{y}_\mathbf{a} ; \boldsymbol\phi)),
\end{equation}}
\fi
\noindent where $f_{\mathbf{Y_a}} (\mathbf{y}_\mathbf{a} ; \boldsymbol\phi)$ is given by~\eqref{eq:jointpdfrewritten} and $\ln f_{\mathbf{Y_a}} (\mathbf{y}_\mathbf{a} ; \boldsymbol\phi)$ is known as the log-likelihood function.

In formulating the channel estimation problem in~\eqref{eq:channelestimationproblem} (and hence in~\eqref{eq:MLE}), we assumed unknown transmitted symbols. These unknown transmitted symbols can be treated as hidden data. A common approach to solving the maximization problem in~\eqref{eq:MLE} in the presence of hidden data is the EM algorithm~\cite{Dempster-1977}, which is adopted in this work. The main steps of EM algorithm are
\begin{enumerate}

\item \textit{Expectation step:} In the $E$-step, the expectation of the log-likelihood is taken over all the values of the hidden variable, conditioned on the vector of observations, and the $n$th estimate of $\boldsymbol{\phi}$ ($\boldsymbol{\phi}^{(n)}$). In~\eqref{eq:nodea}, the hidden variable is $\mathbf{\bar{x}_b}$ and consequently, we need to evaluate $Q(\boldsymbol{\phi}|\boldsymbol{\phi}^{(n)})\triangleq \mathbb{E}_{\mathbf{\bar{X}_b}|\mathbf{y_a},\boldsymbol{\phi}^{(n)}}[\ln{f_{\mathbf{Y_a}}(\mathbf{y_a},\mathbf{\bar{x}_b}|\boldsymbol{\phi})}]$.

\item \textit{Maximization step:} In the $M$-step, the function $Q(\boldsymbol{\phi}|\boldsymbol{\phi}^{(n)})$ obtained from the $E$-step is maximized with respect to $\boldsymbol{\phi}$.

\item \textit{Iterations:} We iterate between the $E$- and $M$-steps until convergence is achieved.
\end{enumerate}

The equations needed for the $E$- and $M$-steps are summarized in the propositions below.

\begin{proposition}\label{prop:em}
The $E$-step during $n$th iteration of the algorithm is given by
\ifCLASSOPTIONpeerreview
\begin{align}\label{eq:Qstep}
Q(\boldsymbol{\phi}|\boldsymbol{\phi}^{(n)}) &=- N\ln(M\pi\sigma^2)-\frac{1}{\sigma^2}\sum_{i=1}^N \sum_{k=1}^M   T_{k,i}^{(n)} |y_{a_i}-h_{ba}\bar{x}_k-h_{aa}\bar{x}_{a_i}|^2,
\end{align}
\else
\begin{align}\label{eq:Qstep}
Q(\boldsymbol{\phi}|&\boldsymbol{\phi}^{(n)}) =- N\ln(M\pi\sigma^2) \nonumber \\ &-\frac{1}{\sigma^2}\sum_{i=1}^N \sum_{k=1}^M   T_{k,i}^{(n)} |y_{a_i}-h_{ba}\bar{x}_k-h_{aa}\bar{x}_{a_i}|^2,
\end{align}
\fi

\noindent where $\boldsymbol{\phi}^{(n)} \triangleq [\hat{h}^{(n)}_{aa}$, $\hat{h}^{(n)}_{ba}]$ are the estimates of the channels obtained from $\boldsymbol{\phi}^{(n)}$ during the $n$th iteration of the algorithm, and $T_{k,i}^{(n)}$ is defined as
\begin{align}\label{eq:T}
 T_{k,i}^{(n)}
&\triangleq \frac{ \exp \left(\frac{-1}{\sigma^2} |y_{a_i}-\hat{h}_{ba}^{(n)} \bar{x}_k-\hat{h}_{aa}^{(n)}\bar{x}_{a_i}|^2 \right)}{\sum_{\bar{k}=1}^M\exp \left(\frac{-1}{\sigma^2} |y_{a_i}-\hat{h}_{ba}^{(n)} \bar{x}_{\bar{k}}-\hat{h}_{aa}^{(n)}\bar{x}_{a_i}|^2  \right)},
\end{align}

\noindent where $k \in \mathcal{K}$, $i \in \mathcal{I}\triangleq [1,2,\cdots,N]$, $\bar{x}_k \in \mathcal{\bar{A}}$, $\bar{x}_{a_i} \in \mathcal{\bar{A}}$ and $\bar{x}_{\bar{k}} \in \mathcal{\bar{A}}$.
\end{proposition}

\begin{IEEEproof}
See Appendix~\ref{appendix: Proposition2}.
\end{IEEEproof}

\begin{proposition}\label{prop:em-m}
The $M$-step during the $n$th iteration of the algorithm is given by
\ifCLASSOPTIONpeerreview
\begin{align} \label{eq:Mstep}
\boldsymbol{\phi}^{(n+1)}& = \frac{1}{s_1s_4-s_2^2-s_3^2}\left[ \begin {array}{c} -s_{{2}}v_{{3}}-s_{{3}}v_{{4}}+s_{{4}}v_{{1}
}\\ \noalign{\medskip}-s_{{2}}v_{{4}}+s_{{3}}v_{{3}}+s_{{4}}v_{{2}}
\\ \noalign{\medskip}s_{{1}}v_{{3}}-s_{{2}}v_{{1}}+s_{{3}}v_{{2}}
\\ \noalign{\medskip}s_{{1}}v_{{4}}-s_{{2}}v_{{2}}-s_{{3}}v_{{1}}
\end {array} \right],
\end{align}
\else
{\small
\begin{align} \label{eq:Mstep}
\boldsymbol{\phi}^{(n+1)}& = \frac{1}{s_1s_4-s_2^2-s_3^2}\left[ \begin {array}{c} -s_{{2}}v_{{3}}-s_{{3}}v_{{4}}+s_{{4}}v_{{1}
}\\ \noalign{\medskip}-s_{{2}}v_{{4}}+s_{{3}}v_{{3}}+s_{{4}}v_{{2}}
\\ \noalign{\medskip}s_{{1}}v_{{3}}-s_{{2}}v_{{1}}+s_{{3}}v_{{2}}
\\ \noalign{\medskip}s_{{1}}v_{{4}}-s_{{2}}v_{{2}}-s_{{3}}v_{{1}}
\end {array} \right],
\end{align}}
\fi
\noindent where
\ifCLASSOPTIONpeerreview
{
\begin{subequations}
\begin{align}
s_1 &\triangleq\sum_{i=1}^N |\bar{x}_{a_i}|^2, \; s_2\triangleq\sum_{i=1}^N \sum_{k=1}^M T_{k,i}^{(n)}\Re(\bar{x}_{a_i} \bar{x}_k^*), \; s_3 \triangleq\sum_{i=1}^N \sum_{k=1}^M T_{k,i}^{(n)}\Im(\bar{x}_{a_i} \bar{x}_k^*), \; s_4\triangleq\sum_{i=1}^N \sum_{k=1}^M T_{k,i}^{(n)} |\bar{x}_k|^2&\label{eq:s1}, \\
v_1&\triangleq\sum_{i=1}^N \Re(\bar{x}^*_{a_i}y_{a_i}), \;\; v_2\triangleq\sum_{i=1}^N \Im(\bar{x}^*_{a_i}y_{a_i}), \;\; v_3\triangleq\sum_{i=1}^N \sum_{k=1}^M T_{k,i}^{(n)}\Re(y_{a_i} \bar{x}_k^*), \;\; v_4\triangleq\sum_{i=1}^N \sum_{k=1}^M T_{k,i}^{(n)}\Im(y_{a_i} \bar{x}_k^*).\label{eq:s4}
\end{align}
\end{subequations}}
\else
{\small
\begin{subequations}
\begin{align}
s_1 &\triangleq\sum_{i=1}^N |x_{a_i}|^2, \;\; s_2\triangleq\sum_{i=1}^N \sum_{k=1}^M T_{k,i}^{(n)}\Re(x_{a_i} \bar{x}_k^*),&\\\label{eq:s1}
s_3 &\triangleq\sum_{i=1}^N \sum_{k=1}^M T_{k,i}^{(n)}\Im(x_{a_i} \bar{x}_k^*), \;\; s_4\triangleq\sum_{i=1}^N \sum_{k=1}^M T_{k,i}^{(n)} |\bar{x}_k|^2,& \\
v_1&\triangleq\sum_{i=1}^N \Re(x^*_{a_i}y_{a_i}), \;\; v_2\triangleq\sum_{i=1}^N \Im(x^*_{a_i}y_{a_i}),&\\
v_3&\triangleq\sum_{i=1}^N \sum_{k=1}^M T_{k,i}^{(n)}\Re(y_{a_i} \bar{x}_k^*), \;\; v_4\triangleq\sum_{i=1}^N \sum_{k=1}^M T_{k,i}^{(n)}\Im(y_{a_i} \bar{x}_k^*).\label{eq:s4}
\end{align}
\end{subequations}}
\fi
\end{proposition}

\begin{IEEEproof}
See Appendix~\ref{appendix: Proposition2}.
\end{IEEEproof}
\begin{remark}
It is well-known that the EM algorithm may be very sensitive to initialization~\cite{Melnykov-2012}. Although different methods exist for EM initialization, generally they are not computationally efficient~\cite{Melnykov-2012,Biernacki-2003}. For the given channel assumptions in Section~\ref{sec:results}, our empirical results showed that initializing the EM algorithm by $\boldsymbol{\phi}^{(0)} \triangleq [0,0,0,0]$ resulted in the lowest estimation error. Hence, this initialization is used in this work.
\end{remark}


\subsection{Lower Bound on the Estimation Error}\label{sec:bounds_analysis}
In this section we derive a closed-form lower bound on the estimation error of the proposed estimator. The derived lower bound directly links the channel estimation error to the parameter $\beta$, defined in Section~\ref{sec:resolving}.

The EM algorithm, defined in Propositions~\ref{prop:em} and~\ref{prop:em-m}, is a ML estimator for the parameter vector $\boldsymbol{\phi}$ in~\eqref{eq:def:phi}. Hence, we aim to derive the lower bound for the variance of the proposed ML estimator. The ML estimator is asymptotically efficient~\cite{Kay-1993} and its MSE is lower bounded by the inverse of the Fisher information matrix (FIM)~\cite{Kay-1993}. This result is known as Cramer Rao lower bound (CRLB) and is given by
\begin{align}\label{eq:CRLB}
\mathbb{E}_{\hat{\Phi}_l}[|\hat{\phi}_l-\phi_l|^2] \geq \left[ \M{I}^{-1}\left[f_{\mathbf{Y_a}}(\mathbf{y}_a;\boldsymbol{\phi})\right]\right]_{l,l},
\end{align}

\noindent where $\phi_{l}$ is the $l$th element of the parameter vector $\boldsymbol{\phi}$, $\hat{\phi}_l$ is an estimate of $\phi_l$, for $l \in \{1,2,3,4\}$, $[\cdot]_{l,l}$ is the $l$th diagonal element of a square matrix, and $\M{I}^{-1}\left[f_{\mathbf{Y_a}}(\mathbf{y}_a;\boldsymbol{\phi})\right]$ is the inverse of FIM. Since the inverse of FIM in \eqref{eq:CRLB} cannot be found in closed-form~\cite{Abdallah-2013,Abdallah-2014}, we derive a lower bound on the MSE of the proposed estimator, which is in closed-from. The result is presented in the proposition below.
\begin{proposition}\label{theorem:crlb}
The variance of the proposed estimator is lower bounded by
\begin{align}\label{eq:bound}
\mathbb{E}_{\hat{\Phi}_l}[|\hat{\phi}_l-\phi_l|^2]\geq \left( \frac{\sigma^2}{2NE} \right)\frac{1+\beta}{(1+2\beta)},
\end{align}

\noindent where $l \in \{1,\cdots,4\}$, $N$ is the number of observations, $E$ is the average symbol energy of the modulation constellation before the shift, and $\beta$ is the portion of $E$ that is allocated to the shift.
\end{proposition}
\begin{IEEEproof}
See Appendix~\ref{sec:bounds}.
\end{IEEEproof}

\begin{remark}
The result in \eqref{eq:bound} links the closed-form lower bound of the estimation error to the average energy of the modulation constellation before the shift and the portion of this average energy allocated to the shift. This is important because in Section~\ref{sec:minimumEnergy}, we will use \eqref{eq:bound} to find the minimum shift energy needed for the proposed technique.
\end{remark}


\subsection{Complexity Analysis}\label{sec:complex}
To evaluate the feasibility in implementing the proposed estimator, we investigate the computational complexity of the estimator in terms of required floating point multiplications and additions (flops)~\cite{Mehrpouyan-2011}.

Table~\ref{Table:Complexity} shows the number of multiplications and additions needed for the EM estimator for $h_{ba}$. Although we only present the complexity analysis of $h_{ba}$, similar complexity is also observed for estimating $h_{aa}$. In each row of the table, the number of required additions and multiplications to implement a given equation is presented and are then summed  to obtain overall complexity.
\ifCLASSOPTIONpeerreview
\begin{table}
\centering
\caption{Complexity analysis of the EM estimator.}
{\small
\begin{tabular}{|c|c|c|}
\hline
\multicolumn{3}{|c|}{EM - Complexity per iteration} \\
\cline{1-3}
 (Eq. No.)   & Additions & Multiplications  \\
\hline
       \eqref{eq:T} &  $3M+2$   & $6M+6$    \\
       \eqref{eq:Qstep} &  $NM(3M+4)$                 & $NM(6M+9)+5$ \\
        \eqref{eq:Mstep} & $4NM(3M+2)+3N+14$ & $4NM(6M+8)+3N+33$ \\
        \hline
        $\hat{h}_{ba}$& $15NM^2+12NM+3N+3M+16$ & $30NM^2+41NM+3N+6M+44$ \\
 \hline
\end{tabular}}
\label{Table:Complexity}
\end{table}
\else
\begin{table}
\centering
\caption{Complexity analysis of the EM estimator.}
\resizebox{0.8\textwidth}{!}{
\begin{tabular}{|c|c|c|}
\hline
\multicolumn{3}{|c|}{EM - Complexity per iteration} \\
\cline{1-3}
 (Eq. No.)   & Additions & Multiplications  \\
\hline
       \eqref{eq:T} &  $3M+2$   & $6M+6$    \\
       \eqref{eq:Qstep} &  $NM(3M+4)$                 & $NM(6M+9)+5$ \\
        \eqref{eq:Mstep} & $4NM(3M+2)+3N+14$ & $4NM(6M+8)+3N+33$ \\
        \hline
        $\hat{h}_{ba}$& $15NM^2+12NM+3N+3M+16$ & $30NM^2+41NM+3N+6M+44$ \\
 \hline
 \end{tabular}}
\label{Table:Complexity}
\end{table}
\fi

It is clear from Table~\ref{Table:Complexity} that the complexity of EM estimator per iteration is proportional to $NM^2$. This analysis shows that the EM algorithm is computationally very efficient since, for a given modulation constellation with size $M$, the computational complexity of the EM estimator only grows linearly with the number of observations, $N$.

\section{Simulation Results}\label{sec:results}

In this section we present numerical and simulation results to investigate the performance of the proposed estimator with asymmetric shifted modulation constellation. We consider a FD communication system as illustrated in Fig.~\ref{fig:SysMod}. The analysis in Section~\ref{sec:bounds_analysis} shows an identical lower bound for the estimation error of both $h_{aa}$ and $h_{ba}$. Hence, in this section, we only present the results for the communication channel $h_{ba}$ since identical results are obtained for the SI channel $h_{aa}$.

For each simulation run, $N$ data and interfering symbols are randomly generated assuming shifted $16$-QAM modulation constellation is used ($M=16$). The channels are constant for the transmission of $N$ symbols, i.e., the quasi static assumption. We assume that there is no line-of-sight (LOS) communication link between the transmitter of node $b$ and the receiver of node $a$. Hence, the communication channel $h_{ba}$ can be modelled as a Rayleigh fading channel, i.e., $h_{ba} \sim \mathcal{CN}(0,\sigma^2_{h_{ba}})$. For the SI channel, experimental results have shown that before passive and active cancellation the SI channel has a strong LOS component and can be modelled as a Rician distribution with a large $K$ factor (approximately 20-25 dB). After passive suppression and analog cancellation, the strong LOS component is significantly reduced but still present and can be modelled as a Ricean distribution with $K=0$ dB~\cite{Duarte-2012}. Hence, we generate the SI channel as
\begin{align}
h_{aa}=\sqrt{\frac{K}{K+1}}\sigma_{h_{aa}}e^{\mathrm{j}\zeta}+\sqrt{\frac{1}{K+1}}\mathcal{CN}(0,\sigma^2_{h_{aa}}),
\end{align}

\noindent where $\zeta$ is uniformly distributed angle of arrival of the LOS component of the SI channel~\cite{Tse-2005}.

For the simulations, the signal-to-interference-noise  ratio (SINR) is given by~\cite{Duarte-2012}
\begin{align}
\text{SINR}=\frac{1}{\frac{1}{\text{SIR}}+\frac{1}{\text{SNR}}},
\end{align}

\noindent where the signal-to-interference ratio $\text{SIR}=\frac{\sigma^2_{h_{ba}}}{\sigma^2_{h_{aa}}}$ assuming both nodes use constellations with the same average energy, the desired-signal-to-noise ratio $\text{SNR}=\frac{\sigma^2_{h_{ba}}\log_2{(M)}E_b}{N_0}$, $E_b$ is the average bit energy which is defined below \eqref{eq:E} and $N_0$ is the noise power spectral density.

As discussed in Section~\ref{sec:intro}, even with state-of-the-art passive suppression and analog cancellation, the SIR can still be around $-50$ dB~\cite{Ahmed-2015,Duarte-2012}. Hence, we adopt this value of the SIR in the simulations while assuming that the communication channel has average energy of unity, i.e., $\mathbb{E}[|h_{ba}|^2]=\sigma^2_{h_{ba}}=1$. Furthermore, in order to investigate the performance of the proposed estimator over a range of SINR, we fix $N_0=1$ and run the simulations for different values of $E_b/N_0$ (in dB). The figures of merit used are the average mean square error (MSE) and the BER, which are obtained by averaging over $5000$ Monte Carlo simulation runs.

\subsection{Minimum Energy Needed for Channel Estimation}\label{sec:minimumEnergy}
In this subsection, we are interested in finding the minimum value of $\beta$, for a given $E_b/N_0$ and $N$. As discussed in Section~\ref{sec:resolving}, we use $s\triangleq \sqrt{\beta E}$, where $0<\beta<1$, to shift the symmetric modulation constellation. Hence, a lower value of $\beta$ is desirable since it means less energy is used to shift the modulation constellation.

In order to find a minimum value of $\beta$ suitable for a practical range of $E_b/N_0$ and $N$, we use the average MSE lower bound in~\eqref{eq:bound} to observe the behavior of the proposed estimator as a function of $\beta$ at low $N$ and low $E_b/N_0$. This is motivated by the fact that the minimum value of $\beta$ found for low $N$ and low $E_b/N_0$ will ensure that the desired estimation error will also be achieved for high $E_b/N_0$ and/or when the number of observations $N$ is large. This intuition is confirmed from~\eqref{eq:bound}, which indicates that higher values of $\beta$ are needed at low $E_b/N_0$ to reach a given estimation error. Furthermore, since the lower bound on the estimation error also decreases with $N$, the minimum value of $\beta$ found for smaller $N$ can also serve for larger $N$. Since the experimental results of~\cite{Duarate-2010,Duarte-2012} show that the FD communication channel is normally constant for more than $N>128$ symbols, we propose to find the minimum $\beta$ at $N=128$ and $E_b/N_0=0$ dB.

\begin{figure}[t]
\centering
\subfigure[MSE performance of the proposed channel estimator vs. $\beta$ for $E_b/N_0=0$ dB, $N=128$ and SIR $= -50$ dB.]
{\label{Fig:minbeta}\scalebox{0.6}{\includegraphics{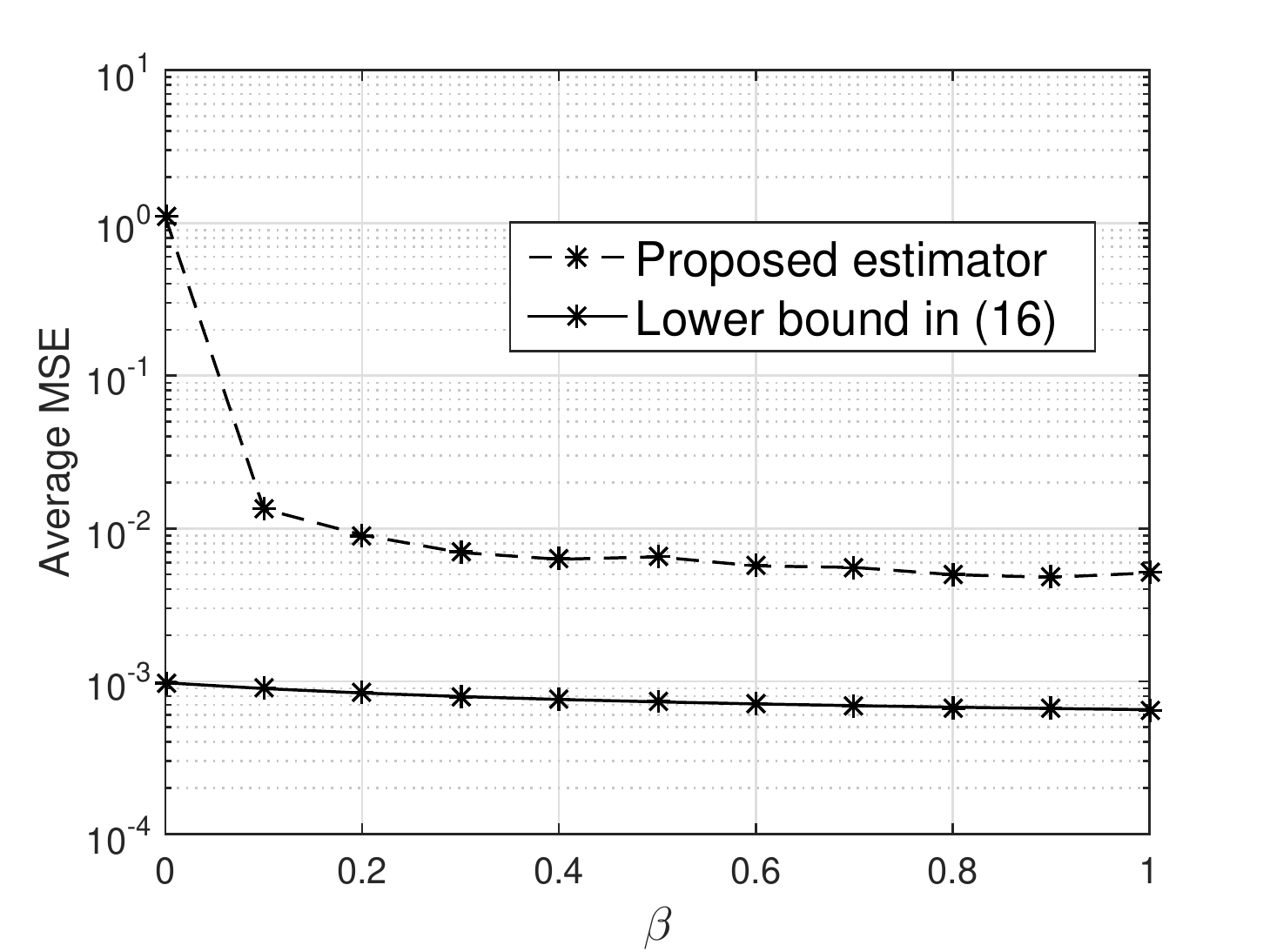}}}%
\subfigure[MSE performance of the proposed channel estimator vs. $E_b/N_0$ for $\beta=0.2$, $N=128$ and SIR $= -50$ dB.]
{\label{Fig:MSE}\scalebox{0.6}{\includegraphics{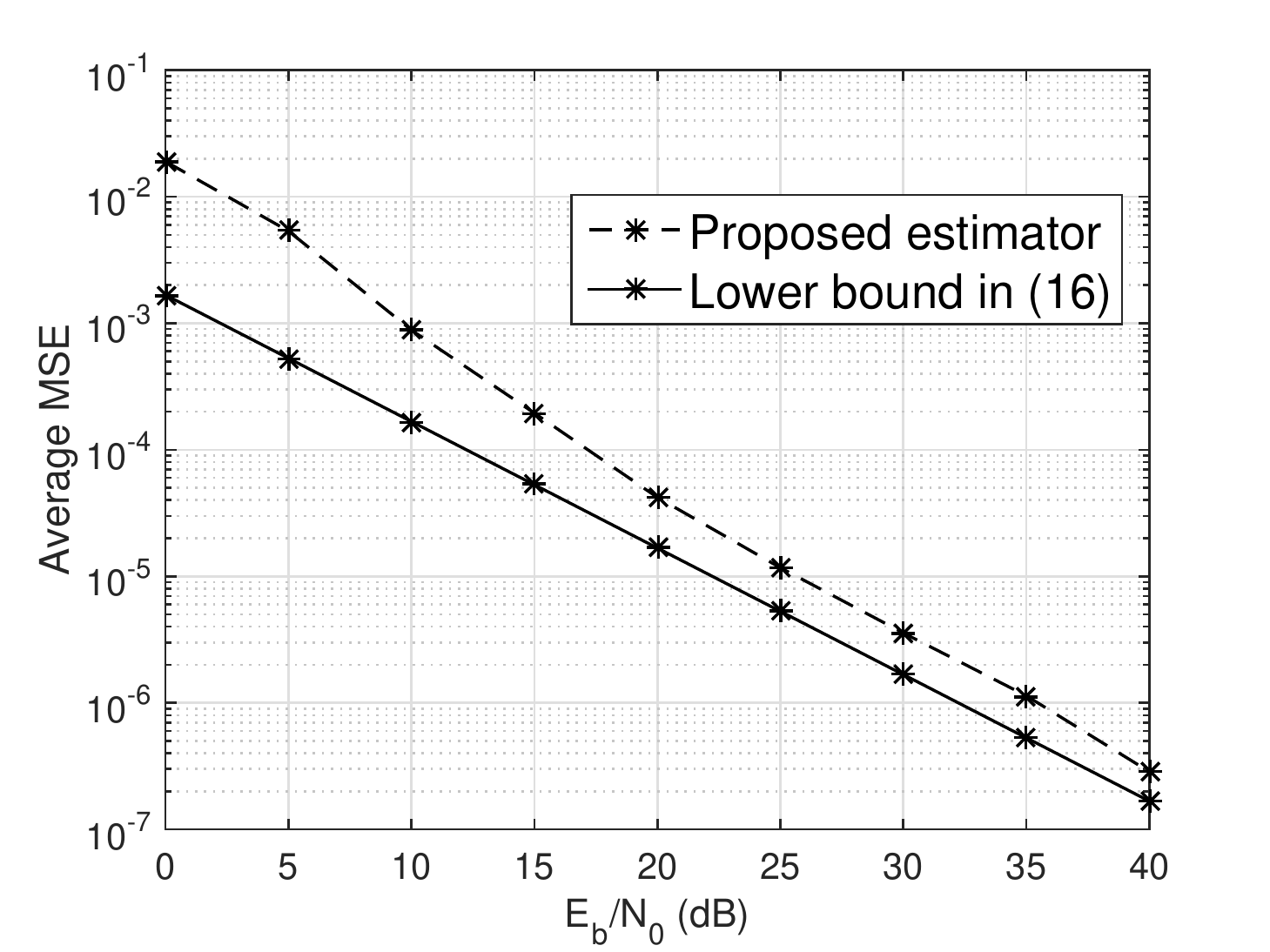}}}
\caption{Performance of proposed estimator with $M=16$-QAM asymmetric shifted modulation constellation.}
\label{Fig:per}
\end{figure}

Fig.~\ref{Fig:minbeta} shows the MSE performance of the proposed technique versus $\beta$ for $E_b/N_0=0$ dB, $N=128$ and SIR $= -50$ dB. If the desired estimation error is taken to be within $10\%$ of the lower bound error, then we can see from the figure that for $\beta <0.2$, the MSE of the proposed estimator is within $10\%$ of the lower bound. Consequently, the minimum value of $\beta$ is $0.2$.

Fig.~\ref{Fig:MSE} shows the MSE performance of the proposed estimator with $\beta=0.2$ (the selected minimum value of $\beta$) vs. $E_b/N_0$ (dB) for $N=128$ and SIR $= -50$ dB. The lower bound in \eqref{eq:bound} is plotted as a reference. The figure shows that as $E_b/N_0$ increases, the gap between the performance of the proposed estimator and the lower bound decreases. The gap is less than $2$ dB after $E_b/N_0=20$ dB.

In the following sections, we set $\beta=0.2$ and $N=128$ to study the performance of the FD communication system.
\subsection{Comparison with Data-Aided Channel Estimation}
\begin{figure}[t!]
  \centering
  \scalebox{0.7}{\includegraphics{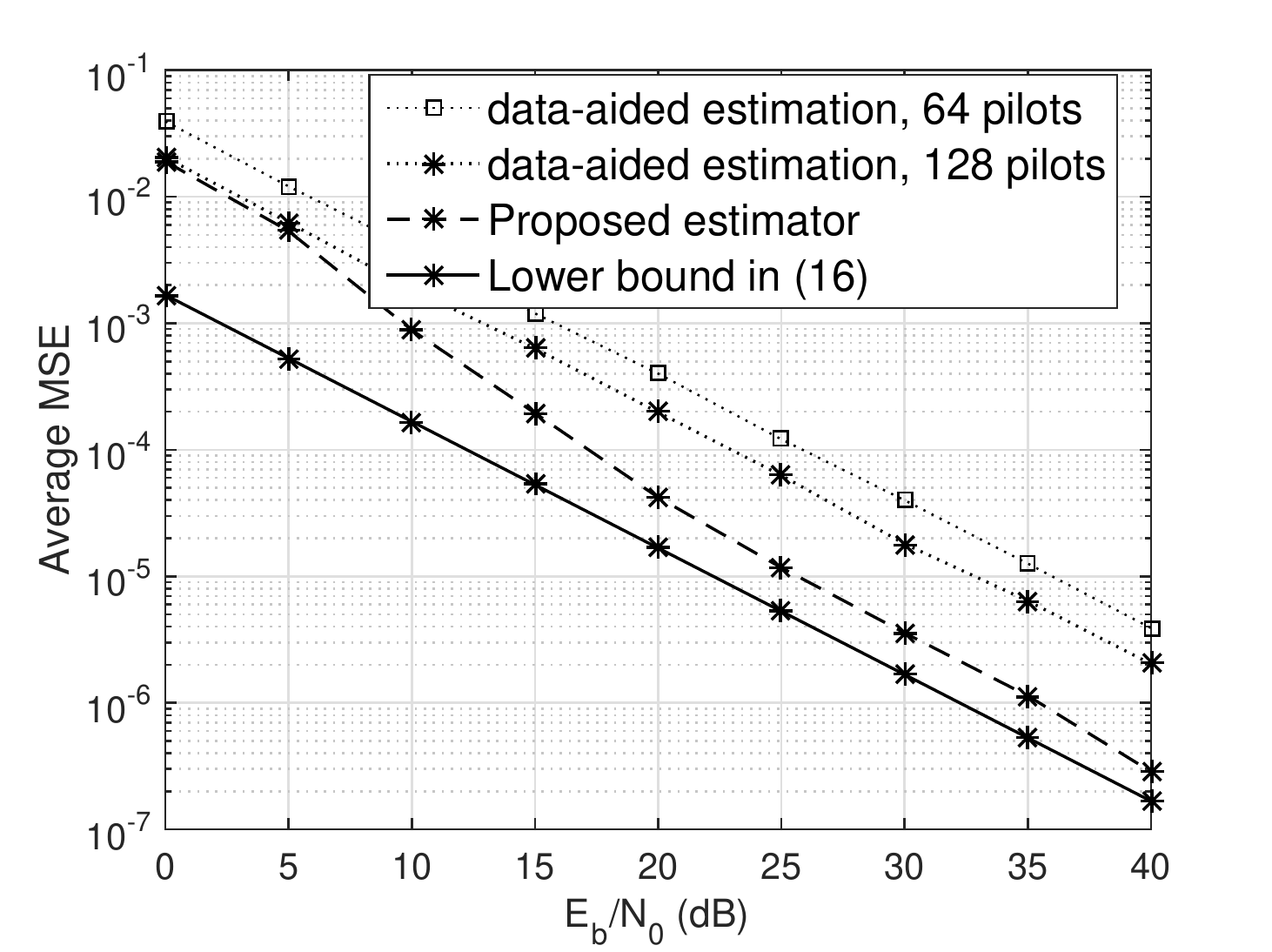}}
  \centering
  \caption{MSE performance of the proposed technique.}\label{fig:PilotMSE}
\end{figure}

In this section we compare the MSE and BER performance of the proposed estimator against a data-aided channel estimator for the case that the average energy per transmitted frame is the same for both methods. For the proposed technique, we assume that (i) all the transmitted symbols are data symbols, and (ii) shifting the modulation constellation increases the average energy by $20\%$ compared to the ideal scenario when no channel estimation is needed (corresponds to $\beta=0.2$). For the data-aided channel estimation, we assume that (i) $64$ pilot symbols are used in a frame of $128$ symbols and (ii) these pilots also require an extra $20\%$ energy.

\underline{\textit{MSE performance:}} The average MSE reveals the accuracy of the channel estimation. Fig.~\ref{fig:PilotMSE} plots the average MSE vs. $E_b/N_0$ with $\beta = 0.2$, $N=128$ and SIR $=-50$ dB. The lower bound from~\eqref{eq:bound} is plotted as a reference. We also plot the MSE for data-aided channel estimation with (i) $64$ pilot symbols in a frame of $128$ symbols and (ii) $128$ pilot symbols in a frame of $128$ symbols. Fig.~\ref{fig:PilotMSE} shows that the proposed technique outperforms data-aided channel estimation when both methods use the same extra amount of energy for channel estimation. At high $E_b/N_0$, the MSE performance of the proposed technique is within $3-4$ dB of the lower bound.
\begin{figure}[t!]
  \centering
  \scalebox{0.7}{\includegraphics{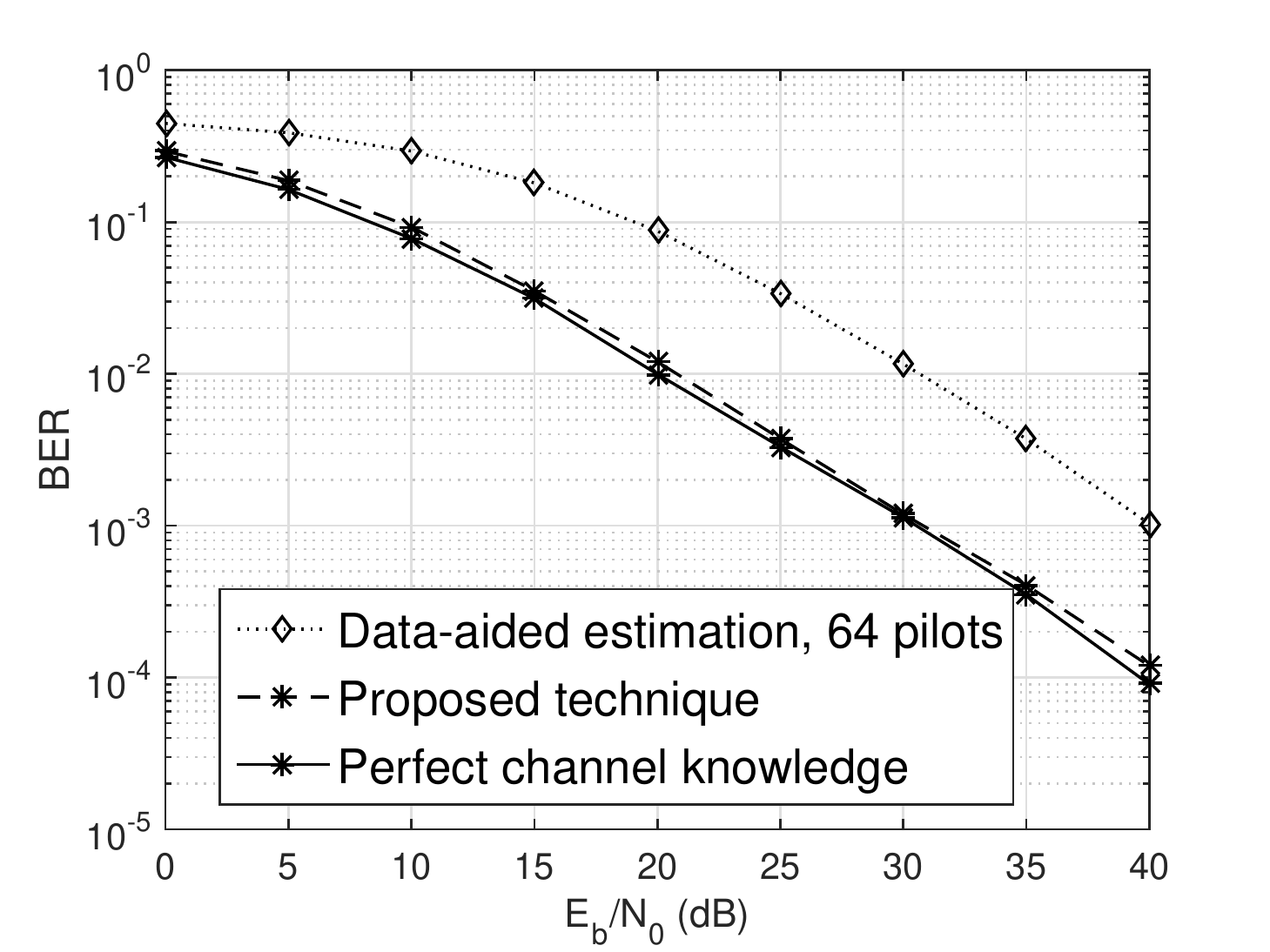}}
  \centering
  \caption{BER performance of the proposed technique.}\label{fig:BER}
\end{figure}

\underline{\textit{BER performance:}} Fig.~\ref{fig:BER} shows the average BER vs. $E_b/N_0$ (dB) with $\beta = 0.2$, $N=128$ and SIR $=-50$ dB. The BER performance with perfect channel knowledge is plotted as a reference. We also plot the BER for data-aided channel estimation with $64$ pilot symbols in a frame of $128$ symbols. Fig.~\ref{fig:BER} shows that the proposed technique outperforms the data-aided channel estimation in terms of the BER. This is to be expected since, as shown in Fig.~\ref{fig:PilotMSE}, for the same extra amount of energy for channel estimation the proposed technique has much lower MSE compared to data-aided channel estimation. In addition, at high $E_b/N_0$, the BER performance of the proposed technique is within $1$ dB of the ideal performance obtained with perfect channel knowledge.

\subsection{Effect of Power of SI Signal}

In the results so far, we have set the SIR to $-50$ dB. In this section, we assess the impact of the SI power level on the performance of the proposed technique.
\begin{figure}[t!]
  \centering
  \scalebox{0.7}{\includegraphics{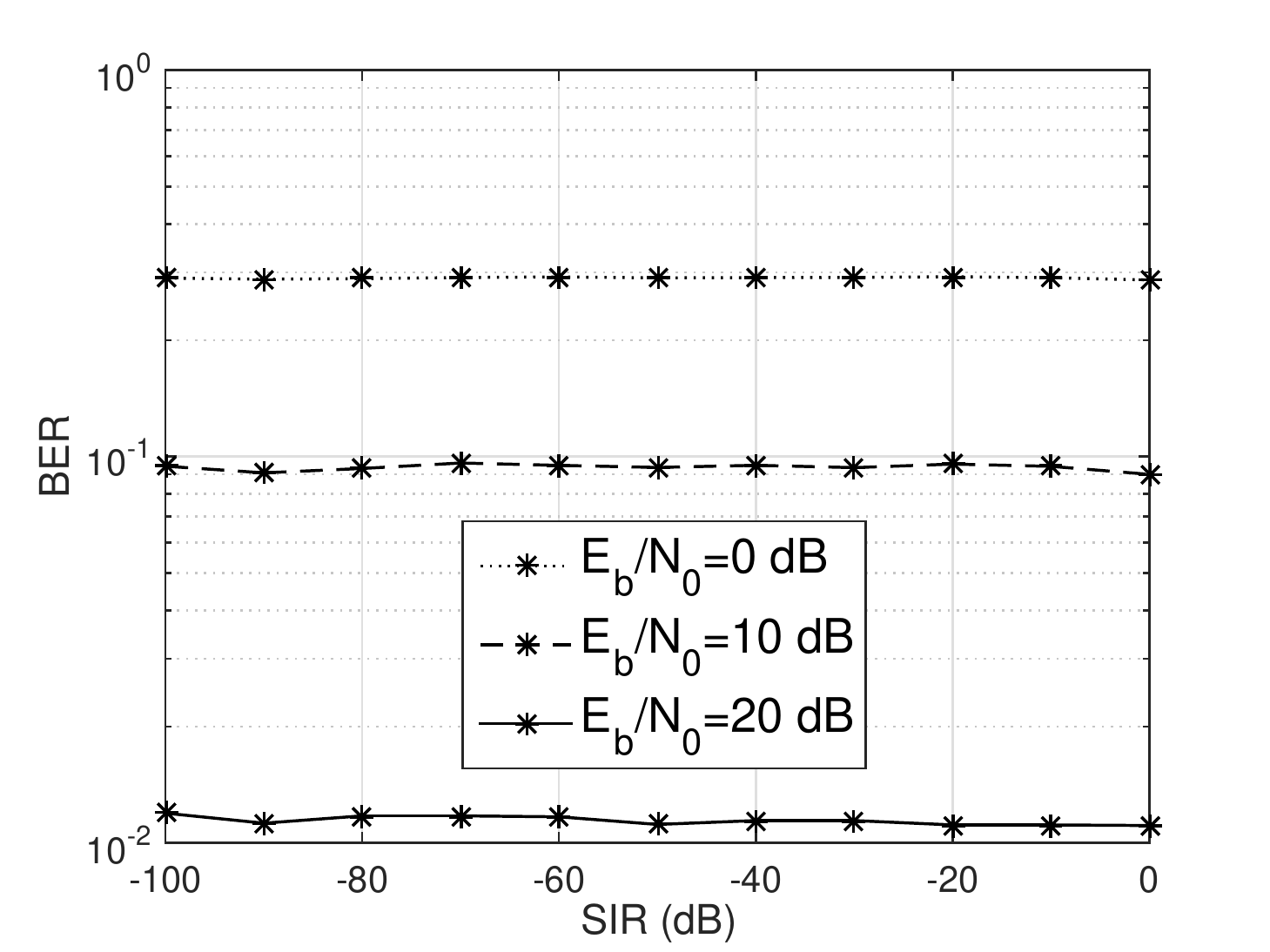}}
  \centering
  \caption{Effect of SI power level on the BER performance of the proposed technique.}\label{fig:SIR}
\end{figure}

Fig.~\ref{fig:SIR} plots the BER versus the SIR (dB) for $E_b/N_0 = 0,10,20$ dB, with $\beta = 0.2$ and $N=128$. \textit{We can see that as the SI power increases, the BER performance of the proposed technique remains nearly constant. This is because in FD communication the self-interference signal is completely known to the receiver~\cite{Sabharwal-2014}. Consequently, for relatively small channel estimation error of the proposed estimator, the SI can be cancelled regardless of its power}. Fig.~\ref{fig:SIR} illustrates the robustness of the proposed technique, i.e., even with weak passive suppression and analog cancellation requiring digital SI cancellation to handle a very large SIR (e.g., $-100$ dB), the BER is not significantly altered.

\section{Conclusions}\label{sec:conc}
In this paper, we have proposed a new technique to estimate the SI and communication channels in a FD communication systems for residual SI cancellation. In the proposed technique, we add a real constant number to each constellation point of a modulation constellation to yield asymmetric shifted modulation constellations with respect to the origin. Using identifiability analysis, we show mathematically that such a modulation constellation can be used for ambiguity-free channel estimation in FD communication systems. We proposed a computationally efficient EM-based estimator to estimate the SI and communication channels simultaneously using the proposed technique. We also derived a lower bound for the estimation error of the proposed estimator. The results showed that the proposed technique is robust to the level of SI power.

\appendices
\numberwithin{equation}{section}
\section{Proof of Theorem~\ref{theorem:identifiability}}\label{appendix:proof:theorem}
The proof consists of three main steps.

\underline{\textit{Step 1:}} We show that if $\boldsymbol{\theta}' \neq \boldsymbol{\theta}$ exists such that $f_{\mathbf{Y_a}}(\mathbf{y_a};\boldsymbol{\theta}')=f_{\mathbf{Y_a}}(\mathbf{y_a};\boldsymbol{\theta})$ $\forall \mathbf{y}_a$, then a bijective function $\mathrm{g}:$ $\mathcal{K}\rightarrow \mathcal{K}$ exists, such that $\frac{x_{k}}{x_{\mathrm{g}(k)}}=c$ $\forall k\in \mathcal{K}$, where $c \neq 1$ is a constant and $|c|=1$. This is done as follows.

If the two joint probability densities $f_{\mathbf{Y_a}}(\mathbf{y_a};\boldsymbol{\theta}')$ and $f_{\mathbf{Y_a}}(\mathbf{y_a};\boldsymbol{\theta})$ are equal $\forall \mathbf{y}_a$, then it easily follows that the marginal densities $f_{Y_{a_i}}(y_{a_i};\boldsymbol{\theta}')$ and $f_{Y_{a_i}}(y_{a_i};\boldsymbol{\theta})$ are also equal $\forall y_{a_i}$. From~\eqref{eq:Marginal}, $f_{Y_{a_i}}(y_{a_i};\boldsymbol{\theta}')$ and $f_{Y_{a_i}}(y_{a_i};\boldsymbol{\theta})$ are given by
    \ifCLASSOPTIONpeerreview
    \begin{subequations}
    \begin{align}\label{appendix:eq:part2:1}
    f_{Y_{a_i}}(y_{a_i};\boldsymbol{\theta}) = \left(\frac{1}{M\pi \sigma^2}\right)\sum_{k=1}^M \exp \left(\frac{-1}{\sigma^2} |y_{a_i}-\boldsymbol{\theta}(1)x_{a_i}-\boldsymbol{\theta}(2)x_{k}|^2\right),
    \end{align}
    \begin{align}\label{appendix:eq:part2:1.b}
    f_{Y_{a_i}}(y_{a_i};\boldsymbol{\theta}')= \left(\frac{1}{M\pi \sigma^2}\right)\sum_{k=1}^M \exp \left(\frac{-1}{\sigma^2} |y_{a_i}-\boldsymbol{\theta}'(1)x_{a_i}-\boldsymbol{\theta}'(2)x_{k}|^2\right).
    \end{align}
    \end{subequations}
    \else
    \begin{subequations}
    \begin{align}\label{appendix:eq:part2:1}
    f_{Y_{a_i}}(y_{a_i};&\boldsymbol{\theta}) = \left(\frac{1}{M\pi \sigma^2}\right)\sum_{k=1}^M&\nonumber\\ &\times\exp \left(\frac{-1}{\sigma^2} |y_{a_i}-\boldsymbol{\theta}(1)x_{a_i}-\boldsymbol{\theta}(2)x_{k}|^2\right)&,
    \end{align}
    \begin{align}\label{appendix:eq:part2:1.b}
    f_{Y_{a_i}}(y_{a_i};&\boldsymbol{\theta}')= \left(\frac{1}{M\pi \sigma^2}\right)\sum_{k=1}^M& \nonumber \\ &{\times\exp \left(\frac{-1}{\sigma^2} |y_{a_i}-\boldsymbol{\theta}'(1)x_{a_i}-\boldsymbol{\theta}'(2)x_{k}|^2\right)&.
    \end{align}
    \end{subequations}
    \fi

If $f_{Y_{a_i}}(y_{a_i};\boldsymbol{\theta}')$ and $f_{Y_{a_i}}(y_{a_i};\boldsymbol{\theta})$ are equal $\forall y_{a_i}$, they should also be equal for $y_{a_i}=\boldsymbol{\theta}(1)x_{a_i}+\boldsymbol{\theta}(2)x_{1}$. In this case, we have
    \begin{align}\label{appendix:eq:part2:3}
    \sum_{k=1}^M \exp \left(\frac{-1}{\sigma^2} |\boldsymbol{\theta}(2)(x_1-x_{k})|^2\right)=\sum_{k=1}^M \exp \left(\frac{-1}{\sigma^2} \left|(\boldsymbol{\theta}(1)-\boldsymbol{\theta}'(1))x_{a_i}+\boldsymbol{\theta}(2)x_1-\boldsymbol{\theta}'(2)x_{k}\right|^2\right).
    \end{align}

The left hand side (LHS) of~\eqref{appendix:eq:part2:3} is independent of $i$, while the right hand side (RHS) of~\eqref{appendix:eq:part2:3} depends on $i$ through $x_{a_i}$. Consequently, for~\eqref{appendix:eq:part2:3} to hold for $\forall y_{a_i}$, the coefficient of $x_{a_i}$ should be zero, i.e., $\boldsymbol{\theta}'(1)=\boldsymbol{\theta}(1)$. Knowing that $\theta'(1)=\theta(1)$ and equating~\eqref{appendix:eq:part2:1} and~\eqref{appendix:eq:part2:1.b}, we have
\begin{align}\label{appendix:eq:part2:4}
\sum_{k=1}^M \exp \left(\frac{-1}{\sigma^2} |y_{a_i}-\boldsymbol{\theta}(1)x_{a_i}-\boldsymbol{\theta}(2)x_{k}|^2\right)=\sum_{k=1}^M \exp \left(\frac{-1}{\sigma^2} |y_{a_i}-\boldsymbol{\theta}(1)x_{a_i}-\boldsymbol{\theta}'(2)x_{k}|^2\right).
\end{align}

By taking the first and second order derivatives of both sides of \eqref{appendix:eq:part2:4} with respect to $y_{a_i}$, it can be shown that $\forall k \in \mathcal{K}$, the points $y_{a_i}= \boldsymbol{\theta}(1)x_{a_i}+\boldsymbol{\theta}(2)x_{k}$ and $y_{a_i}= \boldsymbol{\theta}(1)x_{a_i}+\boldsymbol{\theta}'(2)x_{k}$ maximize the summations of the $M$ exponential functions on the LHS and RHS of~\eqref{appendix:eq:part2:4}, respectively. Consequently, since~\eqref{appendix:eq:part2:4} holds $\forall y_{a_i}$, the points that maximize the summation of $M$ exponential on the LHS of~\eqref{appendix:eq:part2:4} are the same as the points that maximize the summation of $M$ exponentials on the RHS of~\eqref{appendix:eq:part2:4}. Hence, for a bijective function $\mathrm{g}: \mathcal{K}\rightarrow \mathcal{K}$
\begin{align}\label{appendix:eq:part2:5}
x_{a_i}+\boldsymbol{\theta}(2)x_{k}=\boldsymbol{\theta}(1)x_{a_i}+\boldsymbol{\theta}'(2)x_{\mathrm{g}(k)}
\end{align}

It can easily be verified that if~\eqref{appendix:eq:part2:5} holds $\forall y_{a_i}$, then,
\begin{align}\label{appendix:eq:part2:6}
\boldsymbol{\theta}(2)x_k=\boldsymbol{\theta}'(2)x_{\mathrm{g}(k)},
\end{align}

\noindent or
\begin{align}
\frac{\boldsymbol{\theta}'(2)}{\boldsymbol{\theta}(2)}=\frac{x_k}{x_{\mathrm{g}(k)}}.\label{appendix:eq:part2:6.a}
\end{align}
The LHS of~\eqref{appendix:eq:part2:6.a} does not depend on $k$, consequently, the RHS of~\eqref{appendix:eq:part2:6.a} should also be independent of $k$ and should be a constant. Hence, for bijective function $\mathrm{g}$, $\frac{x_k}{x_{\mathrm{g}(k)}}=c$, where $c \neq 1$ is a constant. We note that if $c=1$ then $\theta(2)=\theta'(2)$ and hence $\boldsymbol{\theta}=\boldsymbol{\theta}'$, which violates the assumption that $f_{\mathbf{Y_a}}(\mathbf{y_a};\boldsymbol{\theta}')=f_{\mathbf{Y_a}}(\mathbf{y_a};\boldsymbol{\theta})$ for $\boldsymbol{\theta}\neq \boldsymbol{\theta}'$.

Let us now define permutation $\Pi$ on the ordered set $\mathcal{A}=\{x_1,\cdots,x_M\}$ as
\begin{align}
\Pi\triangleq \left(
  \begin{array}{cccc}
    x_1 & x_2 & \cdots & x_{M} \\
    \Pi(x_1)=x_{\mathrm{g(1)}} &\Pi(x_2)=x_{\mathrm{g(2)}} & \cdots &\Pi(x_M)=x_{\mathrm{g(M)}} \\
  \end{array}
\right).
\end{align}

The sequence $\left(x_k, \Pi(x_k),\Pi(\Pi(x_k)),\cdots, x_k\right)$ forms an orbit of the permutation $\Pi$~\cite{loeher-2011}. If $x_{k}=c x_{\mathrm{g(k)}}$ $\forall k \in \mathcal{K}$, then from the definition of the orbit it is clear that $x_{k}=c^mx_{k}$ $\forall k \in \mathcal{K}$, where $m$ is the length of the orbit sequence and $c\neq 1$ is a constant. Since, $x_k=c^mx_{k}$ $\forall k \in \mathcal{K}$, we can conclude that $c^m=1$ and $|c|=1$.

\underline{\textit{Step 2:}} We show that if the bijective function $\mathrm{g}$: $\mathcal{K} \rightarrow \mathcal{K}$ exits, such that $\frac{x_{k}}{x_{\mathrm{g}(k)}}=c$ $\forall k \in \mathcal{K}$, then there exists a $\boldsymbol{\theta}' \neq \boldsymbol{\theta}$ for which $f_{\mathbf{Y_a}}(\mathbf{y_a};\boldsymbol{\theta}')=f_{\mathbf{Y_a}}(\mathbf{y_a};\boldsymbol{\theta})$ $\forall \mathbf{y}_a$. This is done as follows.

Firstly, in Section~\ref{sec:ambiguity} we showed that the joint PDF of all the observations is given by
\ifCLASSOPTIONpeerreview
\begin{align}\label{appendix:eq:proof:part1:1}
f_{\mathbf{Y_a}}(\mathbf{y_a};\boldsymbol{\theta})=\left(\frac{1}{M\pi \sigma^2}\right)^N \prod_{i=1}^N \sum_{k=1}^M \exp \left(\frac{-1}{\sigma^2} |y_{a_i}-\boldsymbol{\theta}(1)x_{a_i}-\boldsymbol{\theta}(2)x_{k}|^2\right),
\end{align}
\else
\begin{align}\label{appendix:eq:proof:part1:1}
f_{\mathbf{Y_a}}(\mathbf{y_a};&\boldsymbol{\theta}) = \left(\frac{1}{M\pi \sigma^2}\right)^N \prod_{i=1}^N \sum_{k=1}^M&\nonumber\\ &\times\exp \left(\frac{-1}{\sigma^2} |y_{a_i}-\boldsymbol{\theta}(1)x_{a_i}-\boldsymbol{\theta}(2)x_{k}|^2\right)&,
\end{align}
\fi
We can see that for a fixed $i$ in~\eqref{appendix:eq:proof:part1:1}, $M$ different exponential functions corresponding to different values of $x_k \in \mathcal{A}$ are summed together. Since, $\mathrm{g}:\mathcal{K}\rightarrow\mathcal{K}$ is a bijective function and consequently replacing $x_k$ by $x_{\mathrm{g}(x)}$ only affects the order of the exponential functions, we can rewrite~\eqref{appendix:eq:proof:part1:1} as
\ifCLASSOPTIONpeerreview
\begin{align}\label{appendix:eq:proof:part1:2}
f_{\mathbf{Y_a}}(\mathbf{y_a};\boldsymbol{\theta}) = \left(\frac{1}{M\pi \sigma^2}\right)^N \prod_{i=1}^N \sum_{k=1}^M \exp \left(\frac{-1}{\sigma^2} |y_{a_i}-\boldsymbol{\theta}(1)x_{a_i}-\boldsymbol{\theta}(2)x_{\mathrm{g}(k)}|^2\right),
\end{align}
\else
\begin{align}\label{appendix:eq:proof:part1:2}
f_{\mathbf{Y_a}}(\mathbf{y_a};&\boldsymbol{\theta}) = \left(\frac{1}{M\pi \sigma^2}\right)^N \prod_{i=1}^N \sum_{k=1}^M&\nonumber\\ &\times\exp \left(\frac{-1}{\sigma^2} |y_{a_i}-\boldsymbol{\theta}(1)x_{a_i}-\boldsymbol{\theta}(2)x_{\mathrm{g}(k)})|^2\right)&,
\end{align}
\fi

Secondly, for $\boldsymbol{\theta}'\triangleq [\boldsymbol{\theta}(1),\frac{\boldsymbol{\theta}(2)}{c}]$, $f_{\mathbf{Y_a}}(\mathbf{y_a};\boldsymbol{\theta}')$ is given by
\begin{align}\label{appendix:eq:proof:part1:4}
f_{\mathbf{Y_a}}(\mathbf{y_a};\boldsymbol{\theta}')= \left(\frac{1}{M\pi \sigma^2}\right)^N \prod_{i=1}^N \sum_{k=1}^{M} \exp \left(\frac{-1}{\sigma^2} |y_{a_i}-\boldsymbol{\theta}(1)x_{a_i}-\frac{\boldsymbol{\theta}(2)}{c}x_{k}|^2\right),
\end{align}

\noindent and since $\frac{x_k}{c}=x_{\mathrm{g}(k)}$ $\forall k \in \mathcal{K}$,~\eqref{appendix:eq:proof:part1:4} can be written as
\begin{align}\label{appendix:eq:proof:part1:5}
f_{\mathbf{Y_a}}(\mathbf{y_a};\boldsymbol{\theta}')=\left(\frac{1}{M\pi \sigma^2}\right)^N \prod_{i=1}^N \sum_{k=1}^M\exp \left(\frac{-1}{\sigma^2} |y_{a_i}-\boldsymbol{\theta}(1)x_{a_i}-\boldsymbol{\theta}(2)x_{\mathrm{g}(k)}|^2\right),
\end{align}

Comparing~\eqref{appendix:eq:proof:part1:5} with~\eqref{appendix:eq:proof:part1:2} reveals that for $\boldsymbol{\theta}'\neq \boldsymbol{\theta}$, $f_{\mathbf{Y_a}}(\mathbf{y_a};\boldsymbol{\theta}')=f_{\mathbf{Y_a}}(\mathbf{y_a};\boldsymbol{\theta})$ $\forall \mathbf{y}_a$. Consequently, if bijective function $\mathrm{g}$: $\mathcal{K} \rightarrow \mathcal{K}$ exits, such that $\frac{x_{k}}{x_{\mathrm{g}(k)}}=c$ $\forall k \in \mathcal{K}$, then there exists a $\boldsymbol{\theta}' \neq \boldsymbol{\theta}$ for which $f_{\mathbf{Y_a}}(\mathbf{y_a};\boldsymbol{\theta}')=f_{\mathbf{Y_a}}(\mathbf{y_a};\boldsymbol{\theta})$ $\forall \mathbf{y}_a$.

\underline{\textit{Step 3:}} Thirdly, we show that the condition $\frac{x_k}{x_{\mathrm{g}(k)}}=c$  $\forall k \in \mathcal{K}$ is equivalent to the modulation constellation being symmetric around the origin. To prove this equivalency, we need to consider the following two sub-cases:
    \begin{enumerate}[(i)]
    \item Firstly, we need to show that if a bijective function $\mathrm{g}:\mathcal{K}\rightarrow\mathcal{K}$ exists such that $\frac{x_{k}}{x_{\mathrm{g}(k)}}=c$, then the modulation constellation is symmetric with respect to origin. Equivalently, we can show that if a bijective function $\mathrm{g}:\mathcal{K}\rightarrow\mathcal{K}$ does not exist such that $\frac{x_{k}}{x_{\mathrm{g}(k)}}=c$, then the modulation constellation is not symmetric with respect to origin. To prove this equivalent statement, we use proof by contradiction. We assume $\mathrm{g}:\mathcal{K}\rightarrow\mathcal{K}$ does not exist such that $\frac{x_{k}}{x_{\mathrm{g}(k)}}=c$, but the modulation constellation is symmetric with respect to origin. If the modulation is symmetric with respect to the origin then it satisfies the condition of Definition~\ref{def:symm} and hence,
        \begin{align}\label{eq:symmetry}
        \mathrm{f}(-x_k)=-\mathrm{f}(x_k).
        \end{align}

       However, since the function $\mathrm{f}(x_k)$ is defined on set $\mathcal{A}$,~\eqref{eq:symmetry} holds if and only if both $x_k$ and $-x_k$ are in the set $\mathcal{A}$. Consequently, set $\mathcal{A}$ can be represented by
       \begin{align}\label{eq:A:symmetric}
       \mathcal{A}=\{x_1,x_2,\cdots,x_{\frac{M}{2}},-x_1,-x_2,\cdots,-x_{\frac{M}{2}}\}.
       \end{align}

        Now if the bijective function $\mathrm{g}$ is defined as $\mathrm{g}(k)=k+\frac{M}{2}$, then $\frac{x_k}{x_{\mathrm{g}(k)}}=-1$.  However, this contradicts the assumption that bijective function $\mathrm{g}:\mathcal{K}\rightarrow\mathcal{K}$ does not exist, such that $\frac{x_k}{x_{\mathrm{g}(k)}}=c\forall k \in \mathcal{K}$. Hence, if $\mathrm{g}:\mathcal{K}\rightarrow\mathcal{K}$ does not exist such that $\frac{x_{k}}{x_{\mathrm{g}(k)}}=c$, then the modulation constellation cannot be symmetric with respect to origin.

    \item Secondly, we need to show that if the modulation is symmetric then a bijective function $\mathrm{g}:\mathcal{K}\rightarrow \mathcal{K}$ exists such that $\frac{x_k}{x_{\mathrm{g}(k)}}=c$ $\forall k \in \mathcal{K}$. This easily follows from the proof of previous step, where we showed that if the modulation constellation is symmetric then $\mathcal{A}$ can be represented by~\eqref{eq:A:symmetric}. Consequently, a bijective function $\mathrm{g}:\mathcal{K}\rightarrow\mathcal{K}$ exists such that $\frac{x_{k}}{x_{\mathrm{g}(k)}}=-1$ $\forall k \in \mathcal{K}$, i.e., $\mathrm{g}(k)=k+\frac{M}{2}$ $\forall k \in \mathcal{K}$.
    \end{enumerate}

Combining the proofs of the three steps, Theorem~\ref{theorem:identifiability} is proved.

\section{Proof of Propositions \ref{prop:em} and \ref{prop:em-m}}\label{appendix: Proposition2}

Propositions \ref{prop:em} and \ref{prop:em-m} correspond to the $E$ and $M$ steps of the EM algorithm. We assume that both transmitters at nodes $a$ and $b$ use the asymmetric shifted modulation constellation $\mathcal{\bar{A}}$ defined in Definition~\ref{def:asym}, i.e., $\bar{x}_{a_i},\bar{x}_{b_i} \in \mathcal{\bar{A}}$, and assume a uniform discrete distribution for the transmitted symbols.

\subsection{Proof of $E$-Step}\label{appendix:proofEstep}
In the $E$-step of the algorithm function $Q(\boldsymbol{\phi}|\boldsymbol{\phi}^{(n)})$ is given by
\begin{align}\label{appendix:eq:Qstep}
 Q(\boldsymbol{\phi}|\boldsymbol{\phi}^{(n)})=\mathbb{E}_{\mathbf{\bar{X}_b}|\mathbf{y_a},\boldsymbol{\phi}^{(n)}}[\ln{f_{\mathbf{Y_a}}(\mathbf{y_a},\mathbf{\bar{x}_b}|\boldsymbol{\phi})}].
\end{align}

To calculate~\eqref{appendix:eq:Qstep}, we require
$\ln{f_{\mathbf{Y_a}}(\mathbf{y}_a,\mathbf{\bar{x}_b}|\boldsymbol{\phi})}$. Hence, we start with the following joint PDF
%
\ifCLASSOPTIONpeerreview
\begin{align}
  &f_{Y_{a_i}}(y_{a_i},\bar{x}_{b_i};\boldsymbol{\phi})=f_{Y_{a_i}}(y_{a_i}|\bar{X}_{b_i}=\bar{x}_{b_i};\boldsymbol{\phi})p_{\bar{X}_{b_i}}(\bar{x}_{b_i})=\frac{1}{M\pi \sigma^2} \sum_{k =1} ^M  \delta_{\bar{x}_k,\bar{x}_{b_i}} \exp \left(\frac{-1}{\sigma^2} |y_{a_i}-h_{ba} \bar{x}_k-h_{aa}\bar{x}_{a_i}|^2 \right),&
\end{align}
\else
{\small
\begin{align}
  &f_{Y_{a_i}}(y_{a_i},\bar{x}_{b_i};\boldsymbol{\phi})=f_{Y_{a_i}}(y_{a_i}|\bar{X}_{b_i}=\bar{x}_{b_i};\boldsymbol{\phi})p_{\bar{X}_{b_i}}(\bar{x}_{b_i})=\frac{1}{M\pi \sigma^2} \nonumber \\ &\times \sum_{k =1}^M  \delta_{\bar{x}_k,\bar{x}_{b_i}} \exp \left(\frac{-1}{\sigma^2} |y_{a_i}-h_{ba} \bar{x}_{b_i}-h_{aa}\bar{x}_{a_i}|^2 \right),&
\end{align}}
\fi

\noindent where, $\delta_{\bar{x}_k,\bar{x}_{b_i}}$ is the Kronecker delta function and $\delta_{\bar{x}_k,\bar{x}_{b_i}}=1$ if $\bar{x}_{b_i}=\bar{x}_k$ and $0$ otherwise~\cite{Pahl-2001}. Consequently, the log-likelihood of all the observations is given by
\ifCLASSOPTIONpeerreview
\begin{align}\label{appendix:eq:loglik}
\ln( f_{\mathbf{Y_a}}(\mathbf{y_a},\mathbf{x_b};\boldsymbol{\phi}))&= \sum_{i=1}^N \ln( f_{Y_{a_i}}(y_{a_i},\bar{x}_{b_i};\boldsymbol{\phi}))\nonumber\\&=- N\ln(M\pi\sigma^2)-\frac{1}{\sigma^2}\sum_{i=1}^N \sum_{k=1}^M  \delta_{\bar{x}_k,\bar{x}_{b_i}} |y_{a_i}-h_{ba}\bar{x}_{b_i}-h_{aa}\bar{x}_{a_i}|^2.&
\end{align}
\else
{\small
\begin{align}\label{appendix:eq:loglik}
\ln( f_{\mathbf{Y_a}}&(\mathbf{y_a},\mathbf{\bar{x}_b};\boldsymbol{\phi}))= \sum_{i=1}^N \ln( f_{Y_{a_i}}(y_{a_i},\bar{x}_{b_i};\boldsymbol{\phi}))=- N\ln(M\pi\sigma^2)\nonumber \\&-\frac{1}{\sigma^2}\sum_{i=1}^N \sum_{k=1}^M  \delta_{\bar{x}_k,\bar{x}_{b_i}} |y_{a_i}-h_{ba}\bar{x}_{b_i}-h_{aa}\bar{x}_{a_i}|^2.&
\end{align}}
\fi

The expectation in~\eqref{appendix:eq:Qstep} is conditioned on knowing $\boldsymbol{\phi}^{(n)}$ during the $n$th iteration of the algorithm, which is obtained from the $M$-step. Substituting~\eqref{appendix:eq:loglik} in~\eqref{appendix:eq:Qstep}, we have
\ifCLASSOPTIONpeerreview
\begin{align}\label{eq:tempQ}
Q(\boldsymbol{\phi}|\boldsymbol{\phi}^{(n)})&=\mathbb{E}_{\mathbf{\bar{x}_b}|\mathbf{y_a},\boldsymbol{\phi}^{(n)}}[\ln{f_{\mathbf{Y_a}}(\mathbf{y_a},\mathbf{\bar{x}_b};\boldsymbol{\phi})}]\nonumber \\ &=- N\ln(M\pi\sigma^2)-\frac{1}{\sigma^2} \mathbb{E}_{\mathbf{\bar{x}_b}|\mathbf{y_a},\boldsymbol{\phi}^{(n)}}\left[\sum_{i=1}^N \sum_{k=1}^M \delta_{\bar{x}_k,\bar{x}_{b_i}}|y_{a_i}-h_{ba}\bar{x}_{b_i}-h_{aa}\bar{x}_{a_i}|^2\right].&
\end{align}
\else
{\footnotesize
\begin{align}
Q(\boldsymbol{\phi}|\boldsymbol{\phi}^{(n)})=- N\ln(M\pi\sigma^2)-&\frac{1}{\sigma^2} \mathbb{E}_{\mathbf{\bar{x}_b}|\mathbf{y_a},\boldsymbol{\phi}^{(n)}}[\sum_{i=1}^N \sum_{k=1}^M \delta_{\bar{x}_k,\bar{x}_{b_i}}\nonumber \\  &\times|y_{a_i}-h_{ba}\bar{x}_ {b_i}-h_{aa}\bar{x}_{a_i}|^2].&
\end{align}}
\fi

The assumption of independent transmitted symbols allows to rewrite~\eqref{eq:tempQ} as follows
\begin{align}\label{eq:tempQ1}
Q(\boldsymbol{\phi}|\boldsymbol{\phi}^{(n)})&=- N\ln(M\pi\sigma^2)-\frac{1}{\sigma^2}\sum_{i=1}^N \sum_{k=1}^M  \mathbb{E}_{\bar{x}_{b_i}|\mathbf{y_a},\boldsymbol{\phi}^{(n)}}\left[\delta_{\bar{x}_k,\bar{x}_{b_i}}|y_{a_i}-h_{ba}\bar{x}_{b_i}-h_{aa}\bar{x}_{a_i}|^2\right],&\nonumber\\ &=- N\ln(M\pi\sigma^2)-\frac{1}{\sigma^2}\sum_{i=1}^N \sum_{k=1}^M P(\bar{x}_{b_i}=\bar{x}_k|\mathbf{y}_a,\boldsymbol{\phi}^{(n)})|y_{a_i}-h_{ba}\bar{x}_{k}-h_{aa}\bar{x}_{a_i}|^2.&
\end{align}

We define
\begin{align}
T_{k,i}^{(n)} \triangleq P(\bar{x}_{b_i}=\bar{x}_k|\mathbf{y}_a,\boldsymbol{\phi}^{(n)}).
\end{align}

\noindent Then it can easily be shown that
\ifCLASSOPTIONpeerreview
\begin{align}\label{fnT}
 &T_{k,i}^{(n)}=\frac{ \exp \left(\frac{-1}{\sigma^2} |y_{a_i}-\hat{h}_{ba}^{(n)}\bar{ x}_k-\hat{h}_{aa}^{(n)}\bar{x}_{a_i}|^2 \right)}{ \sum_{\bar{k}=1}^M\exp \left(\frac{-1}{\sigma^2} |y_{a_i}-\hat{h}_{ba}^{(n)} \bar{x}_{\bar{k}}-\hat{h}_{aa}^{(n)}\bar{x}_{a_i}|^2 \right)}.&
\end{align}
\else
\begin{align}\label{fnT}\nonumber
 &T_{k,i}^{(n)}=& \nonumber \\
&\frac{ \exp \left(\frac{-1}{\sigma^2} |y_{a_i}-\hat{h}_{ba}^{(n)} \bar{x}_k-\hat{h}_{aa}^{(n)}\bar{x}_{a_i}|^2 \right)}{ \sum_{\bar{k}=1}^M\exp \left(\frac{-1}{\sigma^2} |y_{a_i}-\hat{h}_{ba}^{(n)} \bar{x}_{\bar{k}}-\hat{h}_{aa}^{(n)}\bar{x}_{a_i}|^2 \right) }.&
\end{align}
\fi

Finally, subsituiting~\eqref{fnT} into~\eqref{eq:tempQ1}, $Q(\boldsymbol{\phi}|\boldsymbol{\phi}^{(n)})$ can be found as in~\eqref{appendix:eq:Qstep}. This concludes the proof of Proposition~\ref{prop:em}.
\subsection{Proof of $M$-step}\label{appendix:proofMstep}

The maximization-step of the EM algorithm is given by
\ifCLASSOPTIONpeerreview
\begin{align}\label{appendix:minimization}
\boldsymbol{\phi}^{(n+1)}&=\arg\max_{\boldsymbol{\phi}}Q(\boldsymbol{\phi}|\boldsymbol{\phi}^{(n)})=\arg\min_{\boldsymbol{\phi}} \sum_{i=1}^N \sum_{k=1}^M   T_{k,i}^{(n)} |y_{a_i}-h_{ba}\bar{x}_k-h_{aa}\bar{x}_{a_i}|^2.&
\end{align}
\else
\begin{align}\label{appendix:minimization}
&\boldsymbol{\phi}^{(n+1)}=\arg\max_{\boldsymbol{\phi}}Q(\boldsymbol{\phi}|\boldsymbol{\phi}^{(n)})& \nonumber\\
&=\arg\min_{\boldsymbol{\phi}} \sum_{i=1}^N \sum_{k=1}^M   T_{k,i}^{(n)} |y_{a_i}-\hat{h}^{(n)}_{ba}\bar{x}_k-\hat{h}^{(n)}_{aa}\bar{x}_{a_i}|^2.&
\end{align}
\fi

We define the following function
\begin{align}
\mathrm{r}(\boldsymbol{\phi})\triangleq \sum_{i=1}^N \sum_{k=1}^M   T_{k,i}^{(n)} |y_{a_i}-h_{ba}\bar{x}_k-h_{aa}\bar{x}_{a_i}|^2,&
\end{align}

The minimum of function $\mathrm{\boldsymbol{\phi}}$ (the maximum of the likelihood function), which corresponds to the solution of the $M-$ step of the EM algorithm during the $n$th iteration, happens at the critical point $\boldsymbol{\phi}^{(n+1)}$ for which the Jacobian is zero, i.e., $\M{J}=0$~\cite{Strang-1988}. To find this critical point the Jacobian matrix should be constructed and set equal to zero. This is done by taking the derivative of $\mathrm{r}(\boldsymbol{\phi})$ with respect to the four elements of vector $\boldsymbol{\phi}$, as defined by~\eqref{eq:def:phi}, to construct the Jacobian matrix and then set it equal to zero.  Then, it can be easily shown that the critical point $\boldsymbol{\phi}^{(n+1)}$ is given by
\begin{align}
\boldsymbol{\phi}^{(n+1)}=\M{S}^{-1}\boldsymbol{v},
\end{align}

\noindent where
\begin{align}
\M{S}\triangleq\left[ \begin {array}{cccc} { s_1}&0&{ s_2}&{ s_3}
\\ \noalign{\medskip}0&{ s_1}&-{ s_3}&{ s_2}
\\ \noalign{\medskip}{ s_2}&-{\it s_3}&{ s_4}&0
\\ \noalign{\medskip}{ s_3}&{\ s_2}&0&{ s_4}\end {array} \right]  \quad , \quad \boldsymbol{v}\triangleq \left[ \begin {array}{c} v_{{1}}\\ \noalign{\medskip}v_{{2}}
\\ \noalign{\medskip}v_{{3}}\\\noalign{\medskip}v_{{4}}\end {array}
 \right],
\end{align}

\noindent where the elements of $\M{S} $ and $\boldsymbol{v}$ are given by~\eqref{eq:s1}-\eqref{eq:s4}.

However, to ensure that the critical point $\boldsymbol{\phi}^{(n+1)}$ is the minimum of function $\mathrm{r}(\boldsymbol{\phi})$, the Hessian matrix $\M{H}$ should be positive semi-definite~\cite{Strang-1988}. By taking the second derivatives of $\mathrm{r}(\boldsymbol{\phi})$ with respect to to the four elements of vector $\boldsymbol{\phi}$, we can show that $\M{H}=2\M{S}$. Then, according to Sylvester's criterion~\cite{Strang-1988}, $\M{H}$ is positive semi-definite if and only if all the following are positive
\ifCLASSOPTIONpeerreview
\begin{align}
s_1,\quad \det\left(\left[\begin{array}{cc}
                 s_1 &0 \\
                 0 & s_1
               \end{array}
\right]\right),\quad \det\left(\left[\begin{array}{ccc}
                 s_1 &0&s_2 \\
                 0 & s_1&-s_3\\
                 s_2 & -s_3&s_4
               \end{array}
\right]\right),\quad \det(\M{S}).
\end{align}
\else
\begin{align}
&s_1,\quad \det\left(\left[\begin{array}{cc}
                 s_1 &0 \\
                 0 & s_1
               \end{array}
\right]\right),&\\&\det\left(\left[\begin{array}{ccc}
                 s_1 &0&s_2 \\
                 0 & s_1&-s_3\\
                 s_2 & -s_3&s_4
               \end{array}
\right]\right),\quad \det(\boldsymbol{S}).&
\end{align}
\fi
It can easily be shown that $\det(\M{S})=(s_1s_4-s_2^2-s_2^2)^2$, and is always positive. According to~\eqref{eq:s1}, $s_1$ is always positive, and it is clear that the second determinant is always positive. However, the positivity of the third determinant, i.e., $s_1s_4-s_2^2-s_3^2$, directly depends on the initialization. This is evident from definitions in~\eqref{eq:s1}-\eqref{eq:s4}, which link $s_1$, $s_2$, $s_3$ and $s_4$ to the function $T_{k,i}^{(n)}$ and the derivation of function $T_{k,i}^{(n)}$ in~\eqref{fnT}, which is a function of $\hat{h}^{(n)}_{aa}$ and $\hat{h}^{(n)}_{ba}$, i.e., the estimates from the $n$th iteration. Our numerical investigation shows that for the initialization vector $\boldsymbol{\phi}^{(0)}\triangleq [0,0,0,0]$, the Hessian matrix $\M{H}$ is always positive and hence the critical point $\boldsymbol{\phi}^{(n+1)}$ is indeed the minimum of function $\mathrm{r}(\boldsymbol{\phi})$. Consequently, the EM algorithm with  initialization vector $\boldsymbol{\phi}^{(0)}\triangleq [0,0,0,0]$ converges to the maximum of the likelihood function.

This concludes the proof of Proposition~\ref{prop:em-m}.
\section{Proof of Proposition~\ref{theorem:crlb}}\label{sec:bounds}

It is shown in~\cite{Zamir-1998} that for any random variables $X$ and $Y$ and any parameter $\theta$, if the probability distribution of $X$ is independent of the parameter $\theta$, then $\M{I}[f_{Y}(y;\theta)] < \M{I}[f_{Y}(y|x;\theta)]$, where $\M{I}[\cdot]$ is the FIM and $f(\cdot;\theta)$ is the probability density function parameterized by $\theta$. Consequently, the performance of the proposed estimator is lower bounded by the inverse of $\M{I}[f_{\mathbf{Y_a}}(\mathbf{y_a}|\mathbf{\bar{x}_b};\boldsymbol{\phi})]$, i.e.,
\begin{align}\label{eq:newLB}
\mathbb{E}_{\hat{\Phi}_l}[|\hat{\phi}_l-\phi_l|^2] \geq \left[ \M{I}^{-1}\left[f_{\mathbf{Y_a}}(\mathbf{y_a}|\mathbf{\bar{x}_b};\boldsymbol{\phi})\right]\right]_{l,l} \quad \forall l\in\{1,2,3,4\},
\end{align}
\noindent where $\phi_{l}$ is the $l$th element of the parameter vector $\boldsymbol{\phi}$, $\hat{\phi}_l$ is an estimate of $\phi_l$, and $[\cdot]_{l,l}$ is the $l$th diagonal element of matrix. This is because (i) the performance of the proposed estimator is lower bounded by $\M{I}[f_{\mathbf{Y_a}}(\mathbf{y_a};\boldsymbol{\phi})]$ according to~\eqref{eq:CRLB}, and (ii) $p_{\mathbf{\bar{X}_{b}}}(\mathbf{\bar{x}_{b}})$ is independent of $\boldsymbol{\phi}$. Furthermore, since~\eqref{eq:newLB} holds $\forall \mathbf{\bar{x}_{a}}, \mathbf{\bar{x}_{b}}$, then the variance is also lower-bounded by
\begin{align}\label{eq:newLB1}
\mathbb{E}_{\hat{\Phi}_l}[|\hat{\phi}_l-\phi_l|^2] \geq \left[ \M{I}^{-1}_{avg}\left[f_{\mathbf{Y_a}}(\mathbf{y_a}|\mathbf{\bar{x}_b};\boldsymbol{\phi})\right]\right]_{l,l} \quad \forall l\in\{1,2,3,4\},
\end{align}

\noindent where $\M{I}_{avg}=\mathbb{E}_{\mathbf{\bar{X}_b}, \mathbf{\bar{X}_a}}\left[\M{I}[f_{\mathbf{Y_a}}(\mathbf{y_a}|\mathbf{\bar{x}_b};\boldsymbol{\phi})]\right]$. The value of $\M{I}[f_{\mathbf{Y_a}}(\mathbf{y_a}|\mathbf{\bar{x}_b};\boldsymbol{\phi})]$, needed to evaluate $\M{I}_{avg}$, is presented in the Lemma below.

\begin{lemma} $\M{I}[f(\mathbf{y_a}|\mathbf{\bar{x}_b};\boldsymbol{\phi})]$ is a $4\times 4$ matrix with its elements given by
\ifCLASSOPTIONpeerreview
{\small
\begin{subequations}
\begin{align}
   &i_{1,1}=i_{2,2}=\frac{2}{\sigma^2}\sum_{i=1}^N|\bar{x}_{a_i}|^2, \qquad \mathbb{I}_{3,3}=\mathbb{I}_{4,4}=\frac{2}{\sigma^2}\sum_{i=1}^N|\bar{x}_{b_i}|^2, \\\label{eq:FIM1}
   &i_{2,4}=i_{4,2}=i_{1,3}=i_{3,1}=\frac{2}{\sigma^2}\sum_{i=1}^N \left(\Re\{\bar{x}_{a_i}\}\Re\{\bar{x}_{b_i}\}+\Im\{\bar{x}_{a_i}\}\Im\{\bar{x}_{b_i}\}\right), \\
   &i_{2,3}=i_{3,2}=-i_{1,4}=-i_{4,1}=\frac{2}{\sigma^2}\sum_{i=1}^N \left(\Re\{\bar{x}_{b_i}\}\Im\{\bar{x}_{a_i}\}-\Re\{\bar{x}_{a_i}\}\Im\{\bar{x}_{b_i}\}\right).\label{eq:FIM5}
    \end{align}
\end{subequations}}
\else
{\small
\begin{subequations}
\begin{align}
   i_{1,1}&=i_{2,2}=\frac{2}{\sigma^2}\sum_{i=1}^N|\bar{x}_{a_i}|^2, \\\label{eq:FIM1}
   i_{3,3}&=i_{4,4}=\frac{2}{\sigma^2}\sum_{i=1}^N|\bar{x}_{b_i}|^2, \\
   i_{1,3}&=i_{3,1}=\frac{2}{\sigma^2}\sum_{i=1}^N \left(\Re\{\bar{x}_{a_i}\}\Re\{\bar{x}_{b_i}\}+\Im\{\bar{x}_{a_i}\}\Im\{\bar{x}_{b_i}\}\right), \\
   i_{1,4}&=i_{4,1}=\frac{2}{\sigma^2}\sum_{i=1}^N \left(\Re\{\bar{x}_{a_i}\}\Im\{\bar{x}_{b_i}\}-\Re\{\bar{x}_{b_i}\}\Im\{\bar{x}_{a_i}\}\right), \\
   i_{2,4}&=i_{4,2}=\frac{2}{\sigma^2}\sum_{i=1}^N \left(\Re\{\bar{x}_{a_i}\}\Re\{{x}_{b_i}\}+\Im\{\bar{x}_{a_i}\}\Im\{\bar{x}_{b_i}\}\right), \\
   i_{2,3}&=i_{3,2}=\frac{2}{\sigma^2}\sum_{i=1}^N \left(\Re\{\bar{x}_{b_i}\}\Im\{{x}_{a_i}\}-\Im\{\bar{x}_{b_i}\}\Re\{\bar{x}_{a_i}\}\right).\label{eq:FIM5}
 \end{align}
\end{subequations}}
\fi
\end{lemma}
\begin{IEEEproof}
It can be easily seen from~\eqref{eq:jointpdf1} that $f_{\mathbf{Y_a}}(\mathbf{y_a}|\mathbf{\bar{x}_b};h_{aa},h_{ba})$ is given by
\ifCLASSOPTIONpeerreview
\begin{align}\label{appendix:eq:PDFObservationVector}
&f_{\mathbf{Y_a}}(\mathbf{y_a}|\mathbf{\bar{x}_b};h_{aa},h_{ba})=\left(\frac{1}{\pi\sigma^2}\right)^N\exp\sum_{i=1}^{N}\left(\frac{-|y_{a_i}-h_{aa}\bar{x}_{a_i}-h_{ba}\bar{x}_{b_i}|^2}{\sigma^2}\right).&
\end{align}
\else
\begin{align}\label{appendix:eq:PDFObservationVector}
f_{\mathbf{Y_a}}(\mathbf{y_a}|\mathbf{\bar{x}_b};&h_{aa},h_{ba})=\left(\frac{1}{\pi\sigma^2}\right)^N \nonumber\\&\times\exp\sum_{i=1}^{N}\left(\frac{-|y_{a_i}-h_{aa}\bar{x}_{a_i}-h_{ba}\bar{x}_{b_i}|^2}{\sigma^2}\right).&
\end{align}
\fi
Then, for $l,l'\in\{1,2,3,4\}$, $\M{I}[f_{\mathbf{Y_a}}(\mathbf{y_a}|\mathbf{\bar{x}_b};h_{aa},h_{ba})]$ is~\cite{Kay-1993}
\begin{align}\label{eq:FIM}
  \M{I}_{l,l'}= -\mathbb{E}_{\mathbf{Y_a}}\left[\frac{\partial^2}{\partial_{\phi_m}\partial_{\phi_n}}\ln f_{\mathbf{Y_a}}(\mathbf{y_a}|\mathbf{\bar{x}_b};h_{aa},h_{ba})\right],
\end{align}
\noindent where $m,n \in \{1,2,3,4\}$, $\phi_m$ and $\phi_n$ are the $m$th and $n$th elements of \\ $\boldsymbol{\phi}=[\Re\{h_{aa}\},\Im\{h_{aa}\},\Re\{h_{ba}\},\Im\{h_{ba}\}]$.
By evaluating~\eqref{eq:FIM}, using the joint PDF given by~\eqref{appendix:eq:PDFObservationVector}, the non-zero elements of $\M{I}[f_{\mathbf{Y_a}}(\mathbf{y_a}|\mathbf{\bar{x}_b};h_{aa},h_{ba})]$ can be found and are given by~\eqref{eq:FIM1}-\eqref{eq:FIM5}.
\end{IEEEproof}

Using the value of $\M{I}[f(\mathbf{y_a}|\mathbf{\bar{x}_b};\boldsymbol{\phi})]$ given by the above lemma, we need to evaluate the expectations in order to find $\M{I}_{avg}$. As discussed in Section~\ref{sec:resolving}, we assume that both nodes $a$ and $b$ use a real constant $s\triangleq\sqrt{\beta E}$ to shift the modulation constellation. Since all the constellation points are equally likely to be transmitted, before shifting the modulation constellation we have $\mathbb{E}_{\Im\{\bar{X}_{a_i}\}}[\Im\{x_{a_i}\}]=\mathbb{E}_{\Im\{\bar{x}_{b_i}\}}[\Im\{x_{b_i}\}]=\mathbb{E}_{\Re\{\bar{X}_{a_i}\}}[\Re\{{x}_{a_i}\}]=\mathbb{E}_{\Re\{\bar{X}_{b_i}\}}[\Re\{x_{b_i}\}]=0$. However, after the shift, $\mathbb{E}_{\Re\{\bar{X}_{a_i}\}}[\Re\{\bar{x}_{a_i}\}]=\mathbb{E}_{\Re\{\bar{X}_{b_i}\}}[\Re\{\bar{x}_{b_i}\}]=\sqrt{\beta E}$ and $\mathbb{E}_{\Im\{\bar{X}_{a_i}\}}[\Im\{\bar{x}_{a_i}\}]=\mathbb{E}_{\Im\{\bar{X}_{b_i}\}}[\Im\{\bar{x}_{b_i}\}]=0$. Furthermore, $\mathbb{E}_{X_{a_i}}[|\bar{x}_{a_i}|^2]=\mathbb{E}_{X_{b_i}}[|\bar{x}_{b_i}|^2]=E+\beta E$ since the average energy of the constellation after the shift is increased by the shift energy ($|s|^2=\beta E$). Consequently, the average FIM with respect to $\mathbf{\bar{x}_a}$ and $\mathbf{\bar{x}_b}$ is given by
\begin{align}
\M{I}_{avg}=\frac{2NE}{\sigma^2}\left(
  \begin{array}{cccc}
    1+\beta & 0 & \beta  & 0 \\
    0 &1+\beta & 0 & \beta  \\
    \beta  & 0 &1+\beta & 0 \\
    0 & \beta  & 0 & 1+\beta \\
  \end{array}
\right),
\end{align}
and $\M{I}^{-1}_{avg}$ is given by
\begin{align}\label{eq:tempI}
\M{I}^{-1}_{avg}=\frac{\sigma^2}{2NE}\left(
  \begin{array}{cccc}
    \frac{\beta+1}{(2\beta+1)} & 0 & -\frac{\beta}{(2\beta+1)} & 0 \\
    0 & \frac{\beta+1}{(2\beta+1)}& 0 &-\frac{\beta}{(2\beta+1)}  \\
   -\frac{\beta}{(2\beta+1)} & 0 &\frac{\beta+1}{(2\beta+1)}& 0 \\
    0 & -\frac{\beta}{(2\beta+1)} & 0 & \frac{\beta+1}{(2\beta+1)} \\
  \end{array}\right).
\end{align}

Using~\eqref{eq:newLB1} and considering the diagonal elements of~\eqref{eq:tempI}, we arrive at the result in \eqref{eq:bound}.


\end{document}